\documentclass[%
  reprint,prx,
  showkeys,
  amsfonts,amssymb,amsmath,
  aps,
  nofootinbib,
  dvipsnames
]{revtex4-2}%

\usepackage{comment}
\usepackage{tikz}
\usepackage{subcaption}
\usepackage{graphicx}
\usepackage{enumitem}
\usepackage{physics}
\usepackage{mathtools}
\usepackage{mathrsfs}
\usepackage[hidelinks,colorlinks=true,linkcolor=blue,citecolor=blue]{hyperref}
\usepackage[capitalise,nameinlink]{cleveref}

\makeatletter
\newif\if@supplementary %
\newcommand{\supplementary}{%
  \setcounter{equation}{0}%
  \renewcommand{\theequation}{S\arabic{equation}}%
  \setcounter{figure}{0}%
  \renewcommand{\thefigure}{S\arabic{figure}}%
  \renewcommand{\thesubsection}{\Alph{subsection}}%
  \renewcommand{\thesubsubsection}{\arabic{subsubsection}}%
  \@supplementarytrue%
}
\def\p@subsection{\@ifundefined{@supplementary}{}{\csname if@supplementary\endcsname \fi}}
\def\p@subsubsection{\@ifundefined{@supplementary}{\thesubsection}{\csname if@supplementary\endcsname \thesubsection.\fi}}
\makeatother
\crefname{supp}{Supplementary materials}{Supplementary materials}

\newcommand{\vbm}[1]{\vb*{\mathcal{#1}}}

\newcommand{\rev}[1]{{\color{black}#1}}

\binoppenalty=\maxdimen%
\relpenalty=\maxdimen%

\providecommand{\noopsort}[1]{}

\makeatletter
\def\l@subsubsection#1#2{}
\makeatother

\newlength{\arrow}
\begin{document}

\title{Topological conditions drive stability in meta-ecosystems}

\author{Johannes Nauta}%
\email{johannes.nauta@unipd.it}%
\affiliation{Department of Physics and Astronomy ``Galileo Galilei'', University of Padua, Via F. Marzolo 8, 315126 Padova, Italy}

\author{Manlio De Domenico}%
\affiliation{Department of Physics and Astronomy ``Galileo Galilei'', University of Padua, Via F. Marzolo 8, 315126 Padova, Italy}
\affiliation{Padua Center for Network Medicine, University of Padua, Via F. Marzolo 8, 315126 Padova, Italy}
\affiliation{Istituto Nazionale di Fisica Nucleare, Sez. Padova, Italy}

\date{\today}

\begin{abstract}
  On a global level, ecological communities are being perturbed at an unprecedented rate by human activities and environmental instabilities.
  Yet, we understand little about what factors facilitate or impede long-term persistence of these communities.
  While observational studies indicate that increased biodiversity must, somehow, be driving stability, theoretical studies have argued the exact opposite viewpoint instead.
  This encouraged many researchers to participate in the ongoing \emph{diversity-stability} debate.
  Within this context, however, there has been a severe lack of studies that consider spatial features explicitly, even though nearly all habitats are spatially embedded.
  To this end, we study here the linear stability of meta-ecosystems on networks that describe how discrete patches are connected by dispersal between them.
  By combining results from random-matrix theory and network theory, we are able to show that there are three distinct features that underlie stability: edge density, tendency to triadic closure, and isolation or fragmentation.
  Our results appear to further indicate that network sparsity does not necessarily reduce stability, and that connections between patches are just as, if not more, important to consider when studying the stability of large ecological systems.
\end{abstract}

\maketitle

\section{Introduction}\label{sec:introduction}%
Ecological communities with high diversity and apparent stability are crumbling under global stressors such as rising temperatures and decreasing habitat sizes~\citep{haddad2015habitat,cowie2022sixth}.
These factors are often of human origin and have contributed to a global decline in species diversity~\citep{barnosky2011has,pimm2014biodiversity}.
Yet, while large and diverse ecosystems are ubiquitous, how these systems have assembled and why they are often so resilient is still poorly understood~\citep{jones2009rapid}.
It is therefore vital to understand the mechanisms that enable this apparent resilience, such that these can potentially be put to use to protect endangered ecological communities.

Here, we will focus on unveiling mechanisms that might facilitate stability as resilience to perturbations.
Within this context, in a seminal work May had shown that, under some assumptions, large and complex systems simply cannot be stable~\citep{may1972will}.
By assuming that species interact randomly, May could use methods from \rev{random-matrix theory} to derive a \emph{stability criterion} that determined whether a system would be stable or not.
This gave rise to the well-established \emph{diversity-stability} paradox, igniting debates across distinct scientific communities, from theoretical physics to theoretical ecology, and for which there has not been found a definite answer~\citep{landi2018complexity}.

However, there are some limitations in the random connectivity assumed by May: interactions among species follow more structured patterns~\citep{gross2009generalized,allesina2012stability,allesina2015stability,grilli2016modularity}, are subject to specific constraints~\citep{pettersson2020stability} and in most natural habitats they are spatially extended, meaning that ecosystems are intrinsically patchy or fragmented~\citep{franklin2002what,kefi2007spatial}, with local ecosystems being connected with each other through dispersal or migration~\citep{leibold2004metacommunity}.
When considering patches as nodes and edges as dispersal pathways between the patches~\citep{gilarranz2012spatial}, patchy habitats naturally form a complex network with dynamics both on and between the nodes.

The stabilizing effect of dispersal in such systems appears to rely on environmental fluctuations that are manifested in the heterogeneity of interactions such that they differ significantly between spatially distinct patches~\citep{gravel2016stability,baron2020dispersalinduced,pettersson2021spatial}.
However, note that dispersal may even be destabilizing given the circumstances, especially in combination with trophic structure~\citep{baron2020dispersalinduced}.
Yet, these results have been established without accounting for network structure~\citep{gravel2016stability,baron2020dispersalinduced,garcialorenzana2024interactions}.
Since ecological networks are spatially embedded~\citep{hanski2000metapopulation} and may depend on species-specific dispersal kernels~\citep{viswanathan2008levy,clobert2012dispersal}, the connectivity patterns might even differ depending on the species considered.

\begin{figure}[t]
  \centering
  \includegraphics[width=\columnwidth]{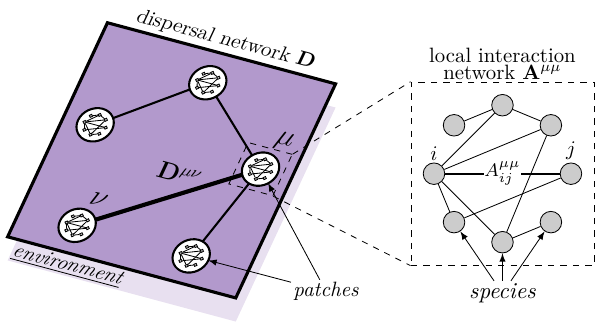}%
  \caption{%
    Illustration of the meta-ecosystems considered in this work (see also~\cref{fig:blockcommunity}).
    \rev{%
      The dispersal network $\vb{D}$, and its matrix entries $\vb{D}^{\mu\nu}$ with elements $D^{\mu\nu}_{ii}$, capture the (rate of) exchange of species $i$ between two connected patches, $\mu$ and $\nu$.
      Within the patches, local interaction networks $\vb{A}^{\mu\mu}$ with elements $A^{\mu\mu}_{ij}$, capture the interaction strengths between species $i$ and $j$.
    }    
  }%
  \label{fig:metaecosystem}
\end{figure}

Here, we focus on the topology of patch-, or dispersal-, networks that comprise a \emph{meta-ecosystem} (\cref{fig:metaecosystem}), and show that connections between patches \rev{significantly influence} the linear stability of ecological systems.
In the following, we shall first introduce our meta-ecosystem model in~\cref{sec:methods} and establish more technical definitions of dispersal and stability in~\cref{sec:dispersal}.
Thereafter, we study several distinct network topologies in~\cref{sec:topology} and discuss our results within the context of ecosystem stability in~\cref{sec:discussion}.
Where applicable, we shall also touch upon possible ventures for experimental verification of our results.

\section{Meta-ecosystems and the community matrix}\label{sec:methods}%
To elucidate the effects of network topology on stability, we adopt an approach similar to that of~\citet{may1972will} and examine the linear stability of a system resting at a hypothetical equilibrium.
We note here that, although it is known that fixed point abundances influence stability (see, e.g., \citep{stone2018feasibility}), our aim is to compare meta-ecosystems with explicit spatial topology to those without.
The effect of (steady state) abundances is thus neglected.

\begin{figure}[b]
  \centering
  \includegraphics[width=.85\columnwidth]{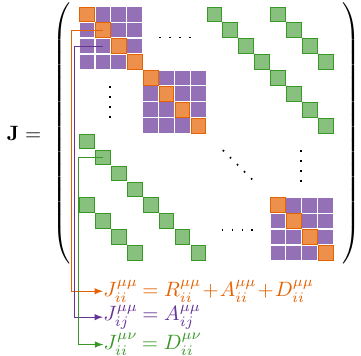}%
  \caption{%
    Block structure of the community matrix $\vb{J}$ of a meta-ecosystem with $S$ species and $M$ patches (see also Refs.~\citep{gravel2016stability,baron2020dispersalinduced}).
    \rev{%
      Diagonal blocks of $\vb{J}$ (orange, purple) --- that is, $\vb{J}^{\mu\mu}$ --- are comprised of a local (diagonal) growth matrix $\vb{R}^{\mu\mu}$, a local interaction matrix $\vb{A}^{\mu\mu}$, and a (diagonal) dispersal matrix $\vb{D}^{\mu\mu}$.
      Off-diagonal blocks (green) are themselves diagonal matrices that capture the species-specific dispersal between patches $\mu$ and $\nu$ (see also~\cref{fig:metaecosystem}).
    }
  }%
  \label{fig:blockcommunity}%
\end{figure}

We consider a meta-ecosystem with $S$ species and $M$ patches (\cref{fig:metaecosystem}).
The ecological dispersal network captures possible dispersal pathways between the $M$ patches.
\rev{Using this network,} our model assumes a community matrix $\vb{J}$ of the form (see, e.g., \citep{gravel2016stability,baron2020dispersalinduced}, \cref{fig:blockcommunity} and \hyperref[sec:supp]{Supplementary materials} for more details)
\begin{align}
  \label{eq:J}
  \vb{J} &= \vb{R} + \vb{A} + \vb{D},
\end{align}
where $\vb{R}$ is a diagonal matrix \rev{representing growth}, $\vb{A}$ the \rev{interaction} matrix, i.e.~the matrix comprised of \rev{local interaction matrices}, and $\vb{D}$ a matrix that defines \rev{(species-specific)} dispersal in between patches.
Within this meta-ecosystem framework, $\vb{J}$ has a block-structure (\cref{fig:blockcommunity}) --- diagonal blocks capture within-patch dynamics while off-diagonal blocks are diagonal matrices that represent (species-specific) between-patch dispersal.

Linear, or asymptotic, stability is governed by the eigenvalues of $\vb{J}$.
More specifically, the criterion for linear stability is that the largest real part of the spectrum is negative (\cref{eq:stabilitycriterion}).
In the absence of dispersal (or when~$M=1$), we recover the standard and well-studied form of the community matrix~\citep{may1972will,allesina2015stability,allesina2020models}.
For a fully-connected network (all-to-all dispersal), it is possible to describe the full spectrum of the community matrix in a closed form, and thus to derive a stability criterion depending on dispersal~\citep{gravel2016stability,baron2020dispersalinduced}.
Conversely, when dispersal occurs on heterogeneous topologies, the fully-connected network model is no longer valid and the spectra of $\vb{J}$ differ significantly depending on both interactions and network topology (see~\cref{fig:classicphaseplots,fig:poissonphaseplots,fig:LRevpoisson,fig:LRevwattsstrogatz,fig:rggphaseplot}), making it difficult to derive closed-form equations for the analysis.
As such, we shall resort here to numerical calculations instead.

Before proceeding to study the influence of network topology on linear stability, let us specify some critical details of our framework.
In the following, we make a distinction between \emph{connected networks}, networks that consist \emph{only} of a single connected component (the giant component), and \emph{disconnected networks}, networks that consist of more than one component.
Within the context of ecological stability this distinction is critical.

This is highlighted by, for example, considering a network that consists of a large, densely connected component that is stable, and a single isolated node.
Within the context of our framework, one could conclude that the addition of a single isolated node renders the system unstable, as some of the eigenvalues corresponding to the isolated patch have positive real part (\cref{fig:SI:isolated}).
However, it is only the single isolated vertex that underlies this instability, and one should question whether a single (isolated and unstable) patch should determine the fate of the entire system.

One can then wonder if such isolated nodes should be considered part of the network or not when studying the system's linear stability.
Here, we assume that permanently isolated patches should not influence macroscopical features such as equilibria and stability, and therefore we resort to studying only networks that are connected --- that is, no isolated nodes exist.
While this assumption is rather strict, we note that the process of becoming isolated is associated with the study of habitat fragmentation \citep{franklin2002what}, which is known to decrease both population abundances and survival probabilities \citep{haddad2015habitat,nicoletti2023emergent}.
Therefore, inclusion of isolated nodes would likely result in unstable systems regardless, and thus would not allow us to study effects of between-patch connections on stability.
A more realistic approach to relax this assumption is to take a multilayer network approach \citep{dedomenico2013mathematical,pilosof2017multilayer,brechtel2018master}, where each layer corresponds to the dispersal network of a specific species.
In this way, isolated patches in one layer might not be isolated in other layers.
However, this more sophisticated framework is beyond the scope of the present work, since our main goal is to understand the role of dispersal in connected ecological patch networks.

\section{Dispersal and stability}\label{sec:dispersal}%
Let us start by considering how dispersal can potentially stabilize ecological systems.
As stated earlier, linear stability is determined by (the sign of) the eigenvalues of the community matrix $\vb{J}$, which depend strongly on dispersal and ecological interaction coefficients (\cref{fig:blockcommunity}).
Let us denote with $\lambda_1 \equiv \lambda_1(\vb{J})$ the \rev{right-most eigenvalue} of $\vb{J}$, i.e.~$\lambda_1 = \max\limits_{i} \Re \lambda_i$, where $\lambda_i \equiv \lambda_i(\vb{J})$ the eigenvalues of $\vb{J}$.
Then, the \emph{stability criterion} reads;
\begin{align}
  \label{eq:stabilitycriterion}
  \Re \lambda_1 < 0
\end{align}

Before proceeding, let us first specify the entries of the community matrix (for more details, see~\cref{sec:SI:communitymatrix}).
For simplicity, we consider the same growth rate $r$ for all species, such that the growth matrix reads
\begin{align}
  \label{eq:growthmatrix}
  \vb{R} = r\vb{I}
\end{align}
Local interaction matrices are assumed to be random matrices, that is
\begin{align}
  \label{eq:interactionmatrix}
  \vb{A} = -b\vb{I} + \vb{B},
\end{align}
where $b$ is the self-interaction term, and $\vb{B}$ is a random block matrix with interactions between species $i$ and $j$ on patches $\mu$ and $\nu$ described by
\begin{align*}
  \langle b_{ij}^{\mu}\rangle = 0, \quad
  \langle {\big(b_{ij}^{\mu}\big)}^2\rangle = c \sigma^2 / S, \quad
  \langle b_{ij}^{\mu}b_{ij}^{\nu}\rangle = \rho c \sigma^2 / S, 
\end{align*}
where we have used the short-hand notation $b_{ij}^\mu \equiv b_{ij}^{\mu\mu}$, as all off-diagonal blocks are $0$ (see also~\cref{fig:blockcommunity}).
The variance includes the \emph{connectance} $c$, which is the probability that elements $b_{ij}^\mu$ are non-zero --- i.e.~the probability that species $i$ and $j$ interact on patch $\mu$ equals $c$.
Spatial heterogeneity is manifested as a correlation between interaction coefficients between two distinct patches $\mu$ and $\nu$ of size $\rho$.
Hence, for $\rho=0$, interactions are i.i.d., and for $\rho=1$ interactions are equal on each patch.
We do not assume negative correlations here.

Finally, we consider homogeneous (diffusive) dispersal with a fixed rate $\gamma$.
Hence, the elements of the dispersal matrix $\vb{D}$ depend on the adjacency matrix $\vbm{G}$ of the patch network as
\begin{align}
  \label{eq:dispersalmatrix}
  D_{ii}^{\mu\nu} =
  \begin{cases}
    -\gamma &\qq{when} \mu=\nu, \\
    \gamma/k_\mu &\qq{when} \mathcal{G}_{\mu\nu}=1,
  \end{cases}
\end{align}
where $k_\mu$ is the degree of patch (node) $\mu$.
Note that with this definition we have
\begin{align*}
  \sum_\nu D^{\mu\nu}_{ii} = 0,
\end{align*}
meaning that dispersal does not bring about potential changes in species abundances.
We additionally assume that dispersal is the same for all species.
For further details on the elements of the community matrix, see~\cref{sec:SI:communitymatrix}.

\begin{figure}[b]
  \centering
  \begin{subcaptiongroup}
    \begin{tikzpicture}
      \subcaptionlistentry{fully-connected graph}\label{phasealltoall}
      \node[inner sep=0pt] (A) at (0,0)
      {\includegraphics[width=.5\columnwidth]{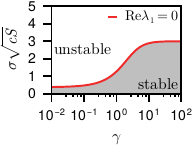}};
      \node[anchor=north west] at ([yshift=1em]A.north west) {\captiontext*{}};      
      \subcaptionlistentry{cyclegraph}\label{phasecycle}
      \node[anchor=west,inner sep=0pt] (C) at (A.east)
      {\includegraphics[width=.5\columnwidth]{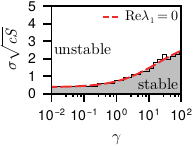}};
      \node[anchor=north west] at ([yshift=1em]C.north west) {\captiontext*{}};
    \end{tikzpicture}
  \end{subcaptiongroup}
  \captionsetup{subrefformat=parens}
  \caption{%
    Phase diagrams of stability versus the complexity, $\sigma\sqrt{cS}$ (see text), and dispersal rate $\gamma$.
    Relevant parameters for both figures are $S=128$, $M=64$, $c=0.2$, $r=1$, $b=1.375$ and $\rho=0.0$.
    Note that $b$ is chosen such that $\sigma\sqrt{cS} = 3$ for $\gamma \gg 1$.
    \subref{phasealltoall}~Phase diagram of fully-connected network.
    Higher dispersal rate allows for higher complexities to remain stable (see~\cref{eq:stabilitycriterionsmallgamma,eq:stabilitycriterionlarge}).
    Red line indicates theoretical boundary at which $\Re\lambda_1=0$ (\cref{sec:SI:esd}).
    \subref{phasecycle}~Phase diagram of cycle-, or ring-network.
    Black lines indicate numerical approximation of the boundary at which $\Re\lambda_1=0$, and dotted red lines are smoothed fitted curves of this boundary.
    Note that the complexity converges for large $\gamma$, but this is difficult to visualize due to numerical instabilities for very large $\gamma$.
  }%
  \label{fig:classicphaseplots}
\end{figure}

In absence of dispersal, i.e.~for $\gamma=0$, all entries of the dispersal matrix are $0$, and we recover the well-established stability criterion that May originally derived~\citep{may1972will,allesina2012stability,allesina2015stability}, which reads
\begin{align}
  \label{eq:Maycriterion}
  \sigma\sqrt{cS} < b - r.
\end{align}
The left hand side of this inequality is often called the \emph{complexity}~\citep{may1972will}.
This criterion arises from \rev{random-matrix theory}, according to which the eigenvalues of a random matrix with mean $r-b$ and variance $c\sigma^2/S$ all lie within a circle with center $(r-b, 0)$ and radius $\sigma\sqrt{cS}$~\citep{tao2010random,allesina2015stability}.
\rev{%
  The complexity and its corresponding stability criterion imply that random systems tend to become unstable the more complex they get.
  Complexity can be adjusted by changing the variance of interactions, the mean number of interactions, or the number of species.
  As a result, it serves as a useful variable and has studied extensively literature since its inception by~\citet{may1972will} (see, e.g., Refs. \citep{allesina2012stability,allesina2020models}).
}%
When dispersal is introduced, the stability criterion changes accordingly. In the case where the patch network is fully connected, the stability criteria have been obtained previously \citep{gravel2016stability,baron2020dispersalinduced}.
For $\gamma$ sufficiently small, the criterion reads
\begin{align}
  \label{eq:stabilitycriterionsmallgamma}
  \sigma\sqrt{cS} < b - r + \gamma, \qquad (\gamma\;\mathrm{small}),
\end{align}
which, again, reduces to May's criterion for $\gamma=0$.
For $\gamma$ sufficiently large it reads instead
\begin{align}
  \label{eq:stabilitycriterionlarge}
  \sigma\sqrt{cS/M} < b - r, \qquad (\gamma\;\mathrm{large}),
\end{align}
which, interestingly, becomes independent of $\gamma$ (\rev{but note that $\gamma$ has to be large}) and depends explicitly on the number of patches.
It is worth mentioning that the criterion of~\cref{eq:stabilitycriterionlarge} can be rewritten by isolating $M$.
In this case, it depends on a minimum system size, i.e.
\begin{align}
  \label{eq:Mmin}
  M > M_{\min} &= \frac{\sigma^2 cS}{{(b-r)}^2}.
\end{align}
This criterion illustrates that, when patch networks are fully connected, there need to be sufficient patches for a system to be stable.
Additionally, the minimum number of patches required for stability increases with the complexity and decreases with increased self-interaction (with respect to the growth rate, i.e., $b > r$).
This indicates that self-regulation is additionally stabilizing, which is in agreement with previous works~\citep{barabas2017selfregulation}.

A phase diagram for the full range of dispersal rates and complexities is shown in~\cref{fig:classicphaseplots}, for both fully connected patch networks and cycle networks.
It illustrates a transition between low and high rates of dispersal, already indicating that how patches are connected affects stability.

\section{Dispersal networks and stability}\label{sec:topology}%
Let us now study the eigenvalues of the community matrix when more realistic and complicated structure is considered.
Recall that we consider only connected networks whose \emph{intraconnectivity} --- i.e.~\emph{how} the nodes are connected --- is defined by their degree distribution.
Data on the degree distributions of ecological patch networks is, rather surprisingly, not readily available.
Despite this, networks are often assumed to exhibit a wide range of degree distributions, ranging from Poisson to (truncated) power-law distributions~\citep{niebuhr2015survival,bertassello2020emergent,nicoletti2023emergent,padmanabha2024spatially}, have modular~\citep{gilarranz2012spatial,gilarranz2020generic} or small-world characteristics~\citep{nicoletti2023emergent,padmanabha2024spatially}, or are explicitly spatially embedded~\citep{grilli2015metapopulation,nicoletti2023emergent}.
However, our results shall indicate that, although the specific topology is important, the stabilizing mechanisms tend to hold across a wide variety of networks.

We initially proceed by specifying the ecological patch network as a configuration model network, i.e.~a network that is generated using the configuration model (see, e.g.~\citep{newman2009random,newman2018networks}), with some degree distribution $p_0(k)$ (see also \cref{sec:SI:configmodel}).
As we are interested in the intraconnectivity of the giant component, we would like to control for the degree distribution corresponding to nodes in the giant component specifically, denoted with $p(k)$.
To this aim, we build on recent studies on articulation points in random networks \citep{tishby2018revealing,tishby2018statistical,tishby2019generating}, which allow us to generate connected networks with any arbitrary degree distribution (see~\cref{sec:SI:configmodel}), thus overcoming the problem related to generating networks that contain isolated patches.

\begin{figure}[b]
  \centering
  \begin{subcaptiongroup}
    \begin{tikzpicture}
      \subcaptionlistentry{poissondispersal}\label{phasepoissondispersal}
      \node[inner sep=0pt] (A) at (0,0)
      {\includegraphics[width=.5\columnwidth]{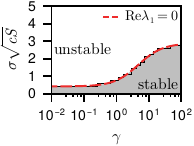}};
      \node[anchor=north west] at ([yshift=1em]A.north west) {\captiontext*{}};      
      \subcaptionlistentry{poissondegree}\label{phasepoissondegree}
      \node[anchor=west,inner sep=0pt] (C) at (A.east)
      {\includegraphics[width=.5\columnwidth]{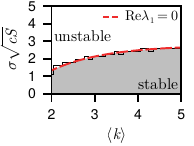}};
      \node[anchor=north west] at ([yshift=1em]C.north west) {\captiontext*{}};
    \end{tikzpicture}
  \end{subcaptiongroup}
  \captionsetup{subrefformat=parens}
  \caption{%
    Phase diagram of stability versus the complexity $\sigma\sqrt{cS}$ (see text) and
    \subref{phasepoissondispersal} the dispersal rate $\gamma$ for $\langle k \rangle=3$, and \subref{phasepoissondegree} the mean degree $\langle k \rangle$ for $\gamma=10$.
    Patch networks are Poisson networks, meaning that they are connected and the (only) connected component has a Poisson degree distribution (\cref{eq:poisson}).
    Systems are maximally spatially heterogeneous with $\rho=0$.
    Black line indicates a numerical approximation of the boundary at which $\Re\lambda_1=0$, and dotted red lines are smoothed fitted curves of this boundary.
    All other parameters are as in~\cref{fig:classicphaseplots}.
  }%
  \label{fig:poissonphaseplots}
\end{figure}

\subsection{Edge density and interaction heterogeneity increase stability}\label{sec:densitystability}%
To investigate the influence of edge density and between-patch heterogeneity, let us initially consider \rev{random connected networks}, named here \emph{Poisson networks}.
That is, we consider networks for which the giant component has a Poisson distribution with minimum degree $k_{\min}=1$ (otherwise isolated patches could exist), which reads
\begin{align}
  \label{eq:poisson}
  p(k; s) = \frac{e^{-s}s^k}{(1-e^{-s}) k!},
\end{align}
where $s = \langle k \rangle$ is the mean degree, and $1-e^{-s}$ is the normalization constant\footnote{Note that the normalization constant differs from the standard Poisson distribution for which $k_{\min}=0$.}.
Numerically obtained phase diagrams for spatially heterogeneous systems ($\rho=0$) with underlying Poisson networks are shown in~\cref{fig:poissonphaseplots}.
They indicate that, as should be expected, the boundary between the stable and unstable regimes lies in between those of fully connected networks (\cref{phasealltoall}) and cycle networks (\cref{phasecycle}).
In addition, increasing the edge density enables systems with higher complexity to remain stable.

\begin{figure}[t]
  \centering
  \includegraphics[width=.85\columnwidth]{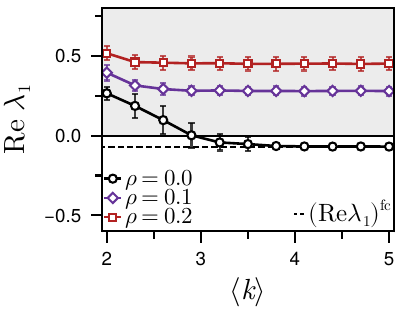}
  \caption{%
    \rev{Real part of the right-most eigenvalue} versus the average degree (edge density) $\langle k \rangle$ of community matrices for which the (only) giant component has a Poisson degree distribution.
    Different interaction heterogeneities $\rho$ are shown.
    Dashed line indicates approximated value in a fully connected network (shown here for $\rho=0$, see also~\cref{sec:SI:esd}).
    Relevant parameters are $S=100$, $M=512$, $c=0.2$, $r=1$, $\sigma\sqrt{cS}=3$, $b=r+\sigma\sqrt{cS/M}$, and $\gamma=10$.
  }%
  \label{fig:LRevpoisson}
\end{figure}

Further inspection of the right-most eigenvalues indicate that edge density --- i.e.,~mean degree $\langle k \rangle$ --- \emph{and} between-patch heterogeneity significantly influence stability (\cref{fig:LRevpoisson}).
Importantly, interaction homogeneity (large~$\rho$) can, under some circumstances, completely prevent a system from becoming stable, no matter how well connected the patches might be.
This point is critical and, while it has been established earlier in fully connected networks and cycle networks~\citep{baron2020dispersalinduced}, our results indicate that this effect might be exaggerated in meta-ecosystems with explicit network structure.

Note that when the edge density is high, the stability criterion approaches the one mentioned earlier, as the patch network becomes closer to a fully connected network.
Interestingly, the edge density that facilitates convergence to the fully connected approximation is not high, especially when compared to the density of a fully connected network $\langle k \rangle_{\textrm{fc}} = M-1$.
This suggests that, although edge density is important for stability, patch networks can be relatively sparse for a stable system to exist.
As long as patches are sufficiently heterogeneous and density is sufficient, stable systems can emerge.

To verify this fact, we have further investigated truly sparse networks (that is, where $\langle k \rangle = O(1)$, see~\cref{sec:SI:sparsenetworks}) and found that the above hypothesis continues to hold in these cases (\cref{fig:SI:LRevExp}).
Therefore, patch networks that could support stable \mbox{(meta-)ecosystems} can be truly sparse.
We note that this result is in line with the growing consensus that real-world networks are generally sparse for reasons rooted on generalized thermodynamics and information exchange~\citep{ghavasieh2024diversity}.
While we do not study here the assembly patterns that govern ecological networks, our results do indicate that sparsity does not necessarily restrict system stability.

Within this context, we would like to touch briefly upon the impact on possible experimental verification of these results.
As recent developments on microcosms allow for a detailed \emph{in vitro} study of microbial \mbox{(meta-)populations} (see, e.g., \citep{venturelli2018deciphering,kurkjian2019metapopulation,larsen2020miniaturizing}, among others), the apparent sparsity could \rev{justify keeping the experimental procedures simple.}
\rev{%
  The reason being that one does not need to include many dispersal pathways to observe stability as if the system was to be fully-connected.
  A possible way one could test this hypothesis is by comparing sparsely connected microcosms, e.g.~using the techniques put forward in Ref.~\citep{kurkjian2019metapopulation}, to fully-connected ones.
  The latter can be obtained by mixing all microcosms during a dilution step.
  We believe that such experiments can greatly enhance our understanding of stability in meta-ecosystems.
}

\subsection{Clustering decreases stability}\label{sec:clustering}%
As the previously discussed topologies do not provide control over vastly different ranges of clustering (that is, global clustering coefficients or triadic closure, see~\cref{sec:SI:smallworld}), we resort here to study stability in small-world networks~\citep{watts1998collective,newman2018networks}.
These networks are constructed by starting with a regular network wherein each node has degree $k$, and each edge is rewired at random with probability $q$ while avoiding self-loops and multiple edges.
Note that in these networks, the edge density --- that is, the total number of edges --- remains fixed once the average degree is fixed, allowing us to study how the structure of the network, and in particular its clustering, influences the linear stability.
\rev{%
  However note that allowing the density to increase does not alter the results significantly (\cref{sec:SI:smallworld:nws}).
}%

\begin{figure}[b]
  \centering
  \includegraphics[width=.85\columnwidth]{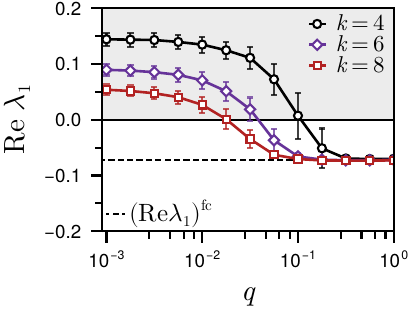}
  \caption{%
    \rev{Real part of the right-most eigenvalue} versus the rewiring probability $q$ of community matrices for which the network is a Watts-Strogatz, or small-world, network.
    Different numbers of initial neighbors before the rewiring procedure, $k$, are shown (see~\cref{sec:SI:smallworld}).
    Note that low $q$ indicates high global clustering coefficients~(\cref{eq:SI:globalclusteringcoefficient}).
    Dashed line indicates approximated value in a fully connected network.
    Relevant parameters are as in~\cref{fig:LRevpoisson}, with $\rho=0$.
  }%
  \label{fig:LRevwattsstrogatz}
\end{figure}

Results for the \rev{real part of the right-most eigenvalues} are shown in~\cref{fig:LRevwattsstrogatz}.
Similar to networks with a Poissonian distribution, stability in small world networks is improved as the density is increased for increased $k$.
However, we can now also appreciate that a high global clustering coefficient (low $q$, see, e.g., \citep{watts1998collective,newman2018networks}) is detrimental towards stability.
This latter result is interesting as opposite viewpoints have been reported previously~\citep{grilli2015metapopulation,nicoletti2023emergent}.
\rev{%
  It is important to stress, however, that the high global clustering coefficient for low $q$ originates from the network being $k$-regular --- that is, the dispersal network is a lattice.
  In metapopulation models, such regular arrangements of patches have been shown to be less stable than more random arrangements~\citep{grilli2015metapopulation}.
  The reason for this is that more random arrangements are more likely to display spatially confined clusters that can act as sources of dispersal~\citep{mouillot2007nicheassembly,gravel2010source,grilli2015metapopulation,loke2023unveiling}, thereby increasing the likelihood of persistence and stability.
}%

However, note that metapopulation models~\citep{hanski1998metapopulation,hanski2000metapopulation}, as opposed to linear models, were considered.
One needs to be careful when comparing metapopulation models with the multi-scale approach of a meta-ecosystems that we consider.
An intuitive reason for this is that metapopulations do not take microscopic processes into account.
This essentially means that some sort of mean-field approach is taken and only the total metapopulation is considered.
In the underlying model presented here, instabilities can, in principle, arise from local interactions.
For example, high levels of clustering do not allow weaker species --- i.e.~those that are generally outcompeted by others --- to easily migrate, hence making their extinction likely and the full system becomes sensitive to (small) perturbation, i.e.~it is unstable.
When the system is instead more homogeneous, the steady state (if it exists) will most likely resemble patterns of niche-partitioning~\citep{wennekes2012neutral}, and might therefore be more likely to be stable.
However, as we consider here only a linearized model, we ourselves should be careful when reasoning about fixed point abundances and their effect on stability in meta-ecosystems and we shall thus refrain from making too strong conclusions.

\rev{%
  Finally, we would like to mention that other network characteristics, such as the mean path length, or perhaps network centralities (and their distributions) may affect system stability.
  We provide a more thorough investigation into the effect of path length on stability in \cref{sec:SI:smallworld}.
  Using an adapted simulated annealing scheme~\citep{reppas2015tuning} to generate dispersal networks with path lengths distinct from the small-world networks reported above, we were able to (slightly) disentangle the effects of clustering and path lengths on stability.
  Our results appear to indicate that clustering more significantly affects stability than path length does, and that path length only becomes a significant factor once relatively low levels of clustering have been reached.
  Despite this, a more in-depth study into underlying network characteristics that influence stability without an analytical motivation is difficult and is considered to be out of the scope of the present analysis.
}

\subsection{Fragmentation-induced instability}\label{sec:randomgeometricgraphs}%
The networks that we have considered up to this point are not geometric networks, meaning that patches are not spatially embedded, and there is no relevant scale associated with the length of the dispersal pathways between patches.
To show that spatially explicit topologies do not drastically change our results, we consider here random geometric graphs~\citep{dall2002random}.
Random geometric graphs are a specific type of spatial networks for which the vertices are distributed in space and edges between them are established only when the (Euclidean) distance between them is lower than some cutoff $\theta$ (\cref{sec:SI:spatialgraphs}). 
When the spatial distribution of vertices is uniform, the networks are usually called random geometric networks, although more complicated or constrained distributions can be considered as well~\citep{herrmann2003connectivity,plaszczynski2022levy}.
The number of patches $M$ and the threshold $\theta$ define the connectivity of the network.
When $\theta > \theta_c^\infty$, a giant component exists, where $\theta_c^\infty$ is the critical threshold for $M\rightarrow \infty$~\citep{dall2002random}.

Random geometric graphs are interesting as they essentially encompass three distinct topological phases: (1)~a phase where most patches are isolated and no giant component exists, for $\theta < \theta_c^\infty$, (2)~a phase where a (sparsely connected) giant component emerges, yet isolated clusters of finite size remain, for $\theta > \theta_c^\infty$, and (3)~a phase where the giant component encompasses the full network and no isolated clusters exist, for $\theta \gg \theta_c^\infty$.

\begin{figure}[t]
  \centering
    \begin{tikzpicture}
      \node[anchor=west,inner sep=0pt] (fig) at (0,0)
      {\includegraphics[width=.7\columnwidth]{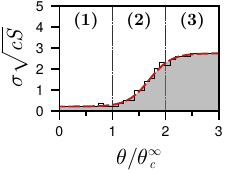}};
      \draw[<-,thick] ([xshift=1.8cm]fig.north west) -- 
      node[above, midway] {\Large fragmentation}
      ([xshift=-.5cm]fig.north east);
    \end{tikzpicture}
  \caption{%
    Fragmentation-induced instability in random geometric graphs.
    Decrease in the (normalized) edge cutoff $\theta/\theta_c^\infty$ can be associated with increased fragmentation.
    Within the phase plot we can identify three distinct regions of destabilizing mechanisms (see~\cref{fig:SI:probcluster}): 
    (1)~fragmentation-induced instability, originating from patches becoming isolated,
    (2)~density-induced instability, originating from edge densities that do not allow for stable systems, and
    (3)~complexity-induced instability, originating from too large and complex systems that cannot be stable, even with high dispersal rates and high network connectivity.
    Here, $S=100$, $M=128$, and all other parameters are as in~\cref{fig:classicphaseplots}.
  }%
  \label{fig:rggphaseplot}
\end{figure}

When studying the phase diagram of stability, we observe that these three phases correspond to three phases of induced instabilities~(\cref{fig:rggphaseplot}).
More specifically, when $\theta<\theta_c^\infty$, patches --- or clusters of patches --- are isolated and are thus subjected to the destabilizing mechanisms of isolation we demonstrated earlier~(\cref{sec:SI:isolated}).
Since the network is spatially embedded, patch isolation is a result of fragmentation, thus the instability that is present here is fragmentation-induced.
When a giant component emerges for $\theta>\theta_c^\infty$ systems with higher complexity are able to remain stable, yet isolated clusters need to be of sufficient size~(see~\cref{eq:Mmin} and \cref{sec:SI:rgg}).
Additionally, edge densities remain low, thus this regime is associated with the density-induced instability that we additionally observed in random and small-world networks.
Finally, when the giant component encompasses the full network for $\theta \gg \theta_c^\infty$, we observe behavior similar as to that in~\cref{fig:poissonphaseplots}, i.e.~a higher edge density typically enables systems with higher complexity to remain stable.
In this regime, the only destabilizing factor is the complexity itself.

In summary, our results are consistent with the idea that fragmentation is a destabilizing mechanism~\citep{grilli2015metapopulation,althagafi2021metapopulation}.
As increased fragmentation rates are being observed globally~\citep{tilman1994habitat,brooks2002habitat}, these results illustrate that maintaining, or increasing, landscape connectivity is most likely key for complex ecosystems to remain stable.

\section{Discussion}\label{sec:discussion}%
We have presented here a study of network-related features, such as the degree distribution, connectance, and clustering coefficients, and their the effects on stability of a linear model.
We considered a diverse set of different network topologies, ranging from random networks to spatial networks,
\rev{%
  where we made use of relatively recent techniques to generate connected networks with arbitrary degree distributions.
}%
In general, regardless of the degree distribution, our results indicate that increases in edge density, corresponding to a sufficient and diverse set of pathways between patches, is imperative for systems to remain stable.
When focusing on networks that display high levels of clustering --- i.e., the tendency to triadic closure --- we found that either high global clustering coefficients or high network regularity (i.e.~similarity to a lattice) contributed negatively to system stability.
Finally, using spatially embedded networks, we highlighted three distinct mechanisms that can induce instabilities; namely fragmentation-induced, edge density-induced, and complexity-induced instability.
Crucially, some of these instabilities cannot be observed in fully-connected systems that previous studies have considered.

While our results are promising, one major shortcoming of our model is the omission of density-dependent effects that materialize during the time evolution of the underlying dynamical model.
Our assumption that the system can be linearized about a feasible steady state is quite strict and should be one of the first things to be relaxed.
However, depending on the interaction structure that is considered, the stability criterion need not necessarily change~\citep{stone2018feasibility}.
In general, the stability criterion depicts an upper bound on the complexity after which a system becomes unstable.
When the interaction structure is more realistic --- e.g.~when extracting interactions from data on food webs --- there appears to still be an upper cutoff on the complexity that still allows for stability~\citep{allesina2012stability,allesina2015stability,grilli2016modularity}.
As the work presented here further establishes an upper bound on the complexity, now depending on the characteristics of the underlying patch network, complexity does not seem to beget stability regardless.

Within this context, while the seminal work of~\citet{may1972will} has spurred debate on the tradeoffs between stability and complexity, recent work has illustrated that increases in complexity might instead be actually stabilizing when sublinear growth rates are considered~\citep{hatton2024diversity}.
Including sublinear growth in the dynamical system at hand drastically changes the eigenvalue distribution, yet it is important to stress that \rev{it is likely that this effect becomes related to population density} %
and will thus not \rev{necessarily} be observed after linearization about the feasible steady state.
\rev{%
  We note, however, that in general one cannot state with certainty that density-dependence changes the stability properties of the system~\citep{stone2018feasibility}.
  However, when introducing non-linear effects in the dynamics, such as extinctions, the surviving communities show distinct characteristics from the initial one~\citep{aguirre-lopez2024heterogeneous,park2024incorporating,poley2024interaction}, and stability measures also become more likely to be effected.
}%
While this again emphasizes the importance of including density-dependence when discussing stability and feasibility of complex ecosystems, what the effects of explicit spatial topologies will be in such systems remains an open problem.

Whereas we have considered the stabilizing effects of dispersal, dispersal can additionally be a destabilizing factor~\citep{hanski1998metapopulation}, as is studied in-depth by~\citet{baron2020dispersalinduced}.
Within the context of meta-ecosystems, this destabilization occurs by virtue of including trophic structure.
That is, a large predator-prey system with distinct average dispersal rates for predator and prey~\citep{baron2020dispersalinduced}.
This introduces activating components, the prey, that are inhibited by others, the predators, giving rise to Turing instabilities --- a phenomenon underlying many dynamics of pattern formation \citep{turing1990chemical}.
However, the potential destabilizing mechanisms of dispersal should be viewed as a separate effect from its stabilizing ones.
The reason is that dispersal induced instability is associated with outliers of the eigenvalue spectrum~\citep{baron2020dispersalinduced}.
In the systems that we have considered, which are, in essence, similar to those that~\citet{gravel2016stability} considered, dispersal affects the bulk of the eigenvalue spectrum and there are no outliers.
However, natural systems are clearly structured~\citep{dunne2002foodweb} and this results in outliers in the spectrum, even when omitting explicit spatial structure \citep{allesina2009food,allesina2015predicting,grilli2016modularity,allesina2020models,poley2023generalized,poley2023eigenvalue}.
Therefore, including trophic structure in meta-ecosystems could elucidate destabilizing effects of dispersal instead, similar to those reported in Ref.~\citep{baron2020dispersalinduced}.

Finally, our results on fragmentation-induced instability further strengthens the fact that more detailed descriptions of complex ecosystems tends to introduce more opportunities for destabilizing mechanisms~\citep{baron2020dispersalinduced}.
Observations on the \mbox{(intra)connectivity} of ecological networks typically shows increased fragmentation rates which effectively decreases edge density, increases patch clustering, and increases the likelihood of subsystems to become isolated~\citep{crooks2011global,crooks2017quantification}.
We have shown here that all these mechanisms are destabilizing.
It would be interesting to experimentally verify these destabilizing mechanisms, for example by employing recent developments in microcosm experiments~\citep{venturelli2018deciphering,kurkjian2019metapopulation,larsen2020miniaturizing}.
Using these developments, one could emulate varying network characteristics by changing dispersal pathways that link distinct wells that house microbial metapopulations.
The fact that our results indicate that networks need not be dense to support stable systems (\cref{fig:poissonphaseplots,fig:LRevpoisson}) may simplify experimental procedures to verify this effect, yet this remains to be seen.
As stability of ecological systems is for now mostly studied theoretically (but see, e.g.,~\citep{yonatan2022complexity,hu2022emergent}), more complex studies of microcosms might reveal potential (de)stabilizing mechanisms such as those studied here.

Overall, our results indicate that if biodiversity is to be maintained, distinct patches will likely need to be ecologically rich and diverse connections between them need to be maintained.
Otherwise, these systems are unlikely to be or remain stable.

\begin{acknowledgments}
  J.N. and M.D.D. acknowledges financial support from the Human Frontier Science Program Organization (HFSP Ref. RGY0064/2022). M.D.D. also acknowledges partial financial support from the INFN grant “LINCOLN” and from MUR funding within the FIS (DD n. 1219 31-07-2023) Project no. FIS00000158.
  The authors would like to thank T. Scagliarini and O. Mazzarisi for insightful comments.
\end{acknowledgments}

\bibliography{bibliography.bib}

\begin{thebibliography}{81}%
\makeatletter
\providecommand \@ifxundefined [1]{%
 \@ifx{#1\undefined}
}%
\providecommand \@ifnum [1]{%
 \ifnum #1\expandafter \@firstoftwo
 \else \expandafter \@secondoftwo
 \fi
}%
\providecommand \@ifx [1]{%
 \ifx #1\expandafter \@firstoftwo
 \else \expandafter \@secondoftwo
 \fi
}%
\providecommand \natexlab [1]{#1}%
\providecommand \enquote  [1]{``#1''}%
\providecommand \bibnamefont  [1]{#1}%
\providecommand \bibfnamefont [1]{#1}%
\providecommand \citenamefont [1]{#1}%
\providecommand \href@noop [0]{\@secondoftwo}%
\providecommand \href [0]{\begingroup \@sanitize@url \@href}%
\providecommand \@href[1]{\@@startlink{#1}\@@href}%
\providecommand \@@href[1]{\endgroup#1\@@endlink}%
\providecommand \@sanitize@url [0]{\catcode `\\12\catcode `\$12\catcode
  `\&12\catcode `\#12\catcode `\^12\catcode `\_12\catcode `\%12\relax}%
\providecommand \@@startlink[1]{}%
\providecommand \@@endlink[0]{}%
\providecommand \url  [0]{\begingroup\@sanitize@url \@url }%
\providecommand \@url [1]{\endgroup\@href {#1}{\urlprefix }}%
\providecommand \urlprefix  [0]{URL }%
\providecommand \Eprint [0]{\href }%
\providecommand \doibase [0]{https://doi.org/}%
\providecommand \selectlanguage [0]{\@gobble}%
\providecommand \bibinfo  [0]{\@secondoftwo}%
\providecommand \bibfield  [0]{\@secondoftwo}%
\providecommand \translation [1]{[#1]}%
\providecommand \BibitemOpen [0]{}%
\providecommand \bibitemStop [0]{}%
\providecommand \bibitemNoStop [0]{.\EOS\space}%
\providecommand \EOS [0]{\spacefactor3000\relax}%
\providecommand \BibitemShut  [1]{\csname bibitem#1\endcsname}%
\let\auto@bib@innerbib\@empty
\bibitem [{\citenamefont {Haddad}\ \emph {et~al.}(2015)\citenamefont {Haddad},
  \citenamefont {Brudvig}, \citenamefont {Clobert}, \citenamefont {Davies},
  \citenamefont {Gonzalez}, \citenamefont {Holt}, \citenamefont {Lovejoy},
  \citenamefont {Sexton}, \citenamefont {Austin}, \citenamefont {Collins},
  \citenamefont {Cook}, \citenamefont {Damschen}, \citenamefont {Ewers},
  \citenamefont {Foster}, \citenamefont {Jenkins}, \citenamefont {King},
  \citenamefont {Laurance}, \citenamefont {Levey}, \citenamefont {Margules},
  \citenamefont {Melbourne}, \citenamefont {Nicholls}, \citenamefont {Orrock},
  \citenamefont {Song},\ and\ \citenamefont {Townshend}}]{haddad2015habitat}%
  \BibitemOpen
  \bibfield  {author} {\bibinfo {author} {\bibfnamefont {N.~M.}\ \bibnamefont
  {Haddad}}, \bibinfo {author} {\bibfnamefont {L.~A.}\ \bibnamefont {Brudvig}},
  \bibinfo {author} {\bibfnamefont {J.}~\bibnamefont {Clobert}}, \bibinfo
  {author} {\bibfnamefont {K.~F.}\ \bibnamefont {Davies}}, \bibinfo {author}
  {\bibfnamefont {A.}~\bibnamefont {Gonzalez}}, \bibinfo {author}
  {\bibfnamefont {R.~D.}\ \bibnamefont {Holt}}, \bibinfo {author}
  {\bibfnamefont {T.~E.}\ \bibnamefont {Lovejoy}}, \bibinfo {author}
  {\bibfnamefont {J.~O.}\ \bibnamefont {Sexton}}, \bibinfo {author}
  {\bibfnamefont {M.~P.}\ \bibnamefont {Austin}}, \bibinfo {author}
  {\bibfnamefont {C.~D.}\ \bibnamefont {Collins}}, \bibinfo {author}
  {\bibfnamefont {W.~M.}\ \bibnamefont {Cook}}, \bibinfo {author}
  {\bibfnamefont {E.~I.}\ \bibnamefont {Damschen}}, \bibinfo {author}
  {\bibfnamefont {R.~M.}\ \bibnamefont {Ewers}}, \bibinfo {author}
  {\bibfnamefont {B.~L.}\ \bibnamefont {Foster}}, \bibinfo {author}
  {\bibfnamefont {C.~N.}\ \bibnamefont {Jenkins}}, \bibinfo {author}
  {\bibfnamefont {A.~J.}\ \bibnamefont {King}}, \bibinfo {author}
  {\bibfnamefont {W.~F.}\ \bibnamefont {Laurance}}, \bibinfo {author}
  {\bibfnamefont {D.~J.}\ \bibnamefont {Levey}}, \bibinfo {author}
  {\bibfnamefont {C.~R.}\ \bibnamefont {Margules}}, \bibinfo {author}
  {\bibfnamefont {B.~A.}\ \bibnamefont {Melbourne}}, \bibinfo {author}
  {\bibfnamefont {A.~O.}\ \bibnamefont {Nicholls}}, \bibinfo {author}
  {\bibfnamefont {J.~L.}\ \bibnamefont {Orrock}}, \bibinfo {author}
  {\bibfnamefont {D.-X.}\ \bibnamefont {Song}},\ and\ \bibinfo {author}
  {\bibfnamefont {J.~R.}\ \bibnamefont {Townshend}},\ }\bibfield  {title}
  {\bibinfo {title} {Habitat fragmentation and its lasting impact on
  {{Earth}}'s ecosystems},\ }\href {https://doi.org/10.1126/sciadv.1500052}
  {\bibfield  {journal} {\bibinfo  {journal} {Science Advances}\ }\textbf
  {\bibinfo {volume} {1}},\ \bibinfo {pages} {e1500052} (\bibinfo {year}
  {2015})}\BibitemShut {NoStop}%
\bibitem [{\citenamefont {Cowie}\ \emph {et~al.}(2022)\citenamefont {Cowie},
  \citenamefont {Bouchet},\ and\ \citenamefont {Fontaine}}]{cowie2022sixth}%
  \BibitemOpen
  \bibfield  {author} {\bibinfo {author} {\bibfnamefont {R.~H.}\ \bibnamefont
  {Cowie}}, \bibinfo {author} {\bibfnamefont {P.}~\bibnamefont {Bouchet}},\
  and\ \bibinfo {author} {\bibfnamefont {B.}~\bibnamefont {Fontaine}},\
  }\bibfield  {title} {\bibinfo {title} {The {{Sixth Mass Extinction}}: Fact,
  fiction or speculation?},\ }\href {https://doi.org/10.1111/brv.12816}
  {\bibfield  {journal} {\bibinfo  {journal} {Biological Reviews}\ }\textbf
  {\bibinfo {volume} {97}},\ \bibinfo {pages} {640} (\bibinfo {year}
  {2022})}\BibitemShut {NoStop}%
\bibitem [{\citenamefont {Barnosky}\ \emph {et~al.}(2011)\citenamefont
  {Barnosky}, \citenamefont {Matzke}, \citenamefont {Tomiya}, \citenamefont
  {Wogan}, \citenamefont {Swartz}, \citenamefont {Quental}, \citenamefont
  {Marshall}, \citenamefont {McGuire}, \citenamefont {Lindsey}, \citenamefont
  {Maguire}, \citenamefont {Mersey},\ and\ \citenamefont
  {Ferrer}}]{barnosky2011has}%
  \BibitemOpen
  \bibfield  {author} {\bibinfo {author} {\bibfnamefont {A.~D.}\ \bibnamefont
  {Barnosky}}, \bibinfo {author} {\bibfnamefont {N.}~\bibnamefont {Matzke}},
  \bibinfo {author} {\bibfnamefont {S.}~\bibnamefont {Tomiya}}, \bibinfo
  {author} {\bibfnamefont {G.~O.~U.}\ \bibnamefont {Wogan}}, \bibinfo {author}
  {\bibfnamefont {B.}~\bibnamefont {Swartz}}, \bibinfo {author} {\bibfnamefont
  {T.~B.}\ \bibnamefont {Quental}}, \bibinfo {author} {\bibfnamefont
  {C.}~\bibnamefont {Marshall}}, \bibinfo {author} {\bibfnamefont {J.~L.}\
  \bibnamefont {McGuire}}, \bibinfo {author} {\bibfnamefont {E.~L.}\
  \bibnamefont {Lindsey}}, \bibinfo {author} {\bibfnamefont {K.~C.}\
  \bibnamefont {Maguire}}, \bibinfo {author} {\bibfnamefont {B.}~\bibnamefont
  {Mersey}},\ and\ \bibinfo {author} {\bibfnamefont {E.~A.}\ \bibnamefont
  {Ferrer}},\ }\bibfield  {title} {\bibinfo {title} {Has the {{Earth}}'s sixth
  mass extinction already arrived?},\ }\href
  {https://doi.org/10.1038/nature09678} {\bibfield  {journal} {\bibinfo
  {journal} {Nature}\ }\textbf {\bibinfo {volume} {471}},\ \bibinfo {pages}
  {51} (\bibinfo {year} {2011})}\BibitemShut {NoStop}%
\bibitem [{\citenamefont {Pimm}\ \emph {et~al.}(2014)\citenamefont {Pimm},
  \citenamefont {Jenkins}, \citenamefont {Abell}, \citenamefont {Brooks},
  \citenamefont {Gittleman}, \citenamefont {Joppa}, \citenamefont {Raven},
  \citenamefont {Roberts},\ and\ \citenamefont
  {Sexton}}]{pimm2014biodiversity}%
  \BibitemOpen
  \bibfield  {author} {\bibinfo {author} {\bibfnamefont {S.~L.}\ \bibnamefont
  {Pimm}}, \bibinfo {author} {\bibfnamefont {C.~N.}\ \bibnamefont {Jenkins}},
  \bibinfo {author} {\bibfnamefont {R.}~\bibnamefont {Abell}}, \bibinfo
  {author} {\bibfnamefont {T.~M.}\ \bibnamefont {Brooks}}, \bibinfo {author}
  {\bibfnamefont {J.~L.}\ \bibnamefont {Gittleman}}, \bibinfo {author}
  {\bibfnamefont {L.~N.}\ \bibnamefont {Joppa}}, \bibinfo {author}
  {\bibfnamefont {P.~H.}\ \bibnamefont {Raven}}, \bibinfo {author}
  {\bibfnamefont {C.~M.}\ \bibnamefont {Roberts}},\ and\ \bibinfo {author}
  {\bibfnamefont {J.~O.}\ \bibnamefont {Sexton}},\ }\bibfield  {title}
  {\bibinfo {title} {The biodiversity of species and their rates of extinction,
  distribution, and protection},\ }\href
  {https://doi.org/10.1126/science.1246752} {\bibfield  {journal} {\bibinfo
  {journal} {Science}\ }\textbf {\bibinfo {volume} {344}},\ \bibinfo {pages}
  {1246752} (\bibinfo {year} {2014})}\BibitemShut {NoStop}%
\bibitem [{\citenamefont {Jones}\ and\ \citenamefont
  {Schmitz}(2009)}]{jones2009rapid}%
  \BibitemOpen
  \bibfield  {author} {\bibinfo {author} {\bibfnamefont {H.~P.}\ \bibnamefont
  {Jones}}\ and\ \bibinfo {author} {\bibfnamefont {O.~J.}\ \bibnamefont
  {Schmitz}},\ }\bibfield  {title} {\bibinfo {title} {Rapid {{Recovery}} of
  {{Damaged Ecosystems}}},\ }\href
  {https://doi.org/10.1371/journal.pone.0005653} {\bibfield  {journal}
  {\bibinfo  {journal} {PLOS ONE}\ }\textbf {\bibinfo {volume} {4}},\ \bibinfo
  {pages} {e5653} (\bibinfo {year} {2009})}\BibitemShut {NoStop}%
\bibitem [{\citenamefont {May}(1972)}]{may1972will}%
  \BibitemOpen
  \bibfield  {author} {\bibinfo {author} {\bibfnamefont {R.~M.}\ \bibnamefont
  {May}},\ }\bibfield  {title} {\bibinfo {title} {Will a {{Large Complex
  System}} be {{Stable}}?},\ }\href {https://doi.org/10.1038/238413a0}
  {\bibfield  {journal} {\bibinfo  {journal} {Nature}\ }\textbf {\bibinfo
  {volume} {238}},\ \bibinfo {pages} {413} (\bibinfo {year}
  {1972})}\BibitemShut {NoStop}%
\bibitem [{\citenamefont {Landi}\ \emph {et~al.}(2018)\citenamefont {Landi},
  \citenamefont {Minoarivelo}, \citenamefont {Br{\"a}nnstr{\"o}m},
  \citenamefont {Hui},\ and\ \citenamefont {Dieckmann}}]{landi2018complexity}%
  \BibitemOpen
  \bibfield  {author} {\bibinfo {author} {\bibfnamefont {P.}~\bibnamefont
  {Landi}}, \bibinfo {author} {\bibfnamefont {H.~O.}\ \bibnamefont
  {Minoarivelo}}, \bibinfo {author} {\bibfnamefont {{\AA}.}~\bibnamefont
  {Br{\"a}nnstr{\"o}m}}, \bibinfo {author} {\bibfnamefont {C.}~\bibnamefont
  {Hui}},\ and\ \bibinfo {author} {\bibfnamefont {U.}~\bibnamefont
  {Dieckmann}},\ }\bibfield  {title} {\bibinfo {title} {Complexity and
  stability of ecological networks: A review of the theory},\ }\href
  {https://doi.org/10.1007/s10144-018-0628-3} {\bibfield  {journal} {\bibinfo
  {journal} {Population Ecology}\ }\textbf {\bibinfo {volume} {60}},\ \bibinfo
  {pages} {319} (\bibinfo {year} {2018})}\BibitemShut {NoStop}%
\bibitem [{\citenamefont {Gross}\ \emph {et~al.}(2009)\citenamefont {Gross},
  \citenamefont {Rudolf}, \citenamefont {Levin},\ and\ \citenamefont
  {Dieckmann}}]{gross2009generalized}%
  \BibitemOpen
  \bibfield  {author} {\bibinfo {author} {\bibfnamefont {T.}~\bibnamefont
  {Gross}}, \bibinfo {author} {\bibfnamefont {L.}~\bibnamefont {Rudolf}},
  \bibinfo {author} {\bibfnamefont {S.~A.}\ \bibnamefont {Levin}},\ and\
  \bibinfo {author} {\bibfnamefont {U.}~\bibnamefont {Dieckmann}},\ }\bibfield
  {title} {\bibinfo {title} {Generalized {{Models Reveal Stabilizing Factors}}
  in {{Food Webs}}},\ }\href {https://doi.org/10.1126/science.1173536}
  {\bibfield  {journal} {\bibinfo  {journal} {Science}\ }\textbf {\bibinfo
  {volume} {325}},\ \bibinfo {pages} {747} (\bibinfo {year}
  {2009})}\BibitemShut {NoStop}%
\bibitem [{\citenamefont {Allesina}\ and\ \citenamefont
  {Tang}(2012)}]{allesina2012stability}%
  \BibitemOpen
  \bibfield  {author} {\bibinfo {author} {\bibfnamefont {S.}~\bibnamefont
  {Allesina}}\ and\ \bibinfo {author} {\bibfnamefont {S.}~\bibnamefont
  {Tang}},\ }\bibfield  {title} {\bibinfo {title} {Stability criteria for
  complex ecosystems},\ }\href {https://doi.org/10.1038/nature10832} {\bibfield
   {journal} {\bibinfo  {journal} {Nature}\ }\textbf {\bibinfo {volume}
  {483}},\ \bibinfo {pages} {205} (\bibinfo {year} {2012})}\BibitemShut
  {NoStop}%
\bibitem [{\citenamefont {Allesina}\ and\ \citenamefont
  {Tang}(2015)}]{allesina2015stability}%
  \BibitemOpen
  \bibfield  {author} {\bibinfo {author} {\bibfnamefont {S.}~\bibnamefont
  {Allesina}}\ and\ \bibinfo {author} {\bibfnamefont {S.}~\bibnamefont
  {Tang}},\ }\bibfield  {title} {\bibinfo {title} {The stability--complexity
  relationship at age 40: A random matrix perspective},\ }\href
  {https://doi.org/10.1007/s10144-014-0471-0} {\bibfield  {journal} {\bibinfo
  {journal} {Popul Ecol}\ }\textbf {\bibinfo {volume} {57}},\ \bibinfo {pages}
  {63} (\bibinfo {year} {2015})}\BibitemShut {NoStop}%
\bibitem [{\citenamefont {Grilli}\ \emph {et~al.}(2016)\citenamefont {Grilli},
  \citenamefont {Rogers},\ and\ \citenamefont
  {Allesina}}]{grilli2016modularity}%
  \BibitemOpen
  \bibfield  {author} {\bibinfo {author} {\bibfnamefont {J.}~\bibnamefont
  {Grilli}}, \bibinfo {author} {\bibfnamefont {T.}~\bibnamefont {Rogers}},\
  and\ \bibinfo {author} {\bibfnamefont {S.}~\bibnamefont {Allesina}},\
  }\bibfield  {title} {\bibinfo {title} {Modularity and stability in ecological
  communities},\ }\href {https://doi.org/10.1038/ncomms12031} {\bibfield
  {journal} {\bibinfo  {journal} {Nat Commun}\ }\textbf {\bibinfo {volume}
  {7}},\ \bibinfo {pages} {12031} (\bibinfo {year} {2016})}\BibitemShut
  {NoStop}%
\bibitem [{\citenamefont {Pettersson}\ \emph {et~al.}(2020)\citenamefont
  {Pettersson}, \citenamefont {Savage},\ and\ \citenamefont
  {Jacobi}}]{pettersson2020stability}%
  \BibitemOpen
  \bibfield  {author} {\bibinfo {author} {\bibfnamefont {S.}~\bibnamefont
  {Pettersson}}, \bibinfo {author} {\bibfnamefont {V.~M.}\ \bibnamefont
  {Savage}},\ and\ \bibinfo {author} {\bibfnamefont {M.~N.}\ \bibnamefont
  {Jacobi}},\ }\bibfield  {title} {\bibinfo {title} {Stability of ecosystems
  enhanced by species-interaction constraints},\ }\href
  {https://doi.org/10.1103/PhysRevE.102.062405} {\bibfield  {journal} {\bibinfo
   {journal} {Phys. Rev. E}\ }\textbf {\bibinfo {volume} {102}},\ \bibinfo
  {pages} {062405} (\bibinfo {year} {2020})}\BibitemShut {NoStop}%
\bibitem [{\citenamefont {Franklin}\ \emph {et~al.}(2002)\citenamefont
  {Franklin}, \citenamefont {Noon},\ and\ \citenamefont
  {George}}]{franklin2002what}%
  \BibitemOpen
  \bibfield  {author} {\bibinfo {author} {\bibfnamefont {A.~B.}\ \bibnamefont
  {Franklin}}, \bibinfo {author} {\bibfnamefont {B.~R.}\ \bibnamefont {Noon}},\
  and\ \bibinfo {author} {\bibfnamefont {T.~L.}\ \bibnamefont {George}},\
  }\bibfield  {title} {\bibinfo {title} {What is habitat fragmentation?},\
  }\href@noop {} {\bibfield  {journal} {\bibinfo  {journal} {Studies in avian
  biology}\ }\textbf {\bibinfo {volume} {25}},\ \bibinfo {pages} {20} (\bibinfo
  {year} {2002})}\BibitemShut {NoStop}%
\bibitem [{\citenamefont {K{\'e}fi}\ \emph {et~al.}(2007)\citenamefont
  {K{\'e}fi}, \citenamefont {Rietkerk}, \citenamefont {Alados}, \citenamefont
  {Pueyo}, \citenamefont {Papanastasis}, \citenamefont {ElAich},\ and\
  \citenamefont {{\noopsort{ruiter}}{de Ruiter}}}]{kefi2007spatial}%
  \BibitemOpen
  \bibfield  {author} {\bibinfo {author} {\bibfnamefont {S.}~\bibnamefont
  {K{\'e}fi}}, \bibinfo {author} {\bibfnamefont {M.}~\bibnamefont {Rietkerk}},
  \bibinfo {author} {\bibfnamefont {C.~L.}\ \bibnamefont {Alados}}, \bibinfo
  {author} {\bibfnamefont {Y.}~\bibnamefont {Pueyo}}, \bibinfo {author}
  {\bibfnamefont {V.~P.}\ \bibnamefont {Papanastasis}}, \bibinfo {author}
  {\bibfnamefont {A.}~\bibnamefont {ElAich}},\ and\ \bibinfo {author}
  {\bibfnamefont {P.~C.}\ \bibnamefont {{\noopsort{ruiter}}{de Ruiter}}},\
  }\bibfield  {title} {\bibinfo {title} {Spatial vegetation patterns and
  imminent desertification in {{Mediterranean}} arid ecosystems},\ }\href
  {https://doi.org/10.1038/nature06111} {\bibfield  {journal} {\bibinfo
  {journal} {Nature}\ }\textbf {\bibinfo {volume} {449}},\ \bibinfo {pages}
  {213} (\bibinfo {year} {2007})}\BibitemShut {NoStop}%
\bibitem [{\citenamefont {Leibold}\ \emph {et~al.}(2004)\citenamefont
  {Leibold}, \citenamefont {Holyoak}, \citenamefont {Mouquet}, \citenamefont
  {Amarasekare}, \citenamefont {Chase}, \citenamefont {Hoopes}, \citenamefont
  {Holt}, \citenamefont {Shurin}, \citenamefont {Law}, \citenamefont {Tilman},
  \citenamefont {Loreau},\ and\ \citenamefont
  {Gonzalez}}]{leibold2004metacommunity}%
  \BibitemOpen
  \bibfield  {author} {\bibinfo {author} {\bibfnamefont {M.~A.}\ \bibnamefont
  {Leibold}}, \bibinfo {author} {\bibfnamefont {M.}~\bibnamefont {Holyoak}},
  \bibinfo {author} {\bibfnamefont {N.}~\bibnamefont {Mouquet}}, \bibinfo
  {author} {\bibfnamefont {P.}~\bibnamefont {Amarasekare}}, \bibinfo {author}
  {\bibfnamefont {J.~M.}\ \bibnamefont {Chase}}, \bibinfo {author}
  {\bibfnamefont {M.~F.}\ \bibnamefont {Hoopes}}, \bibinfo {author}
  {\bibfnamefont {R.~D.}\ \bibnamefont {Holt}}, \bibinfo {author}
  {\bibfnamefont {J.~B.}\ \bibnamefont {Shurin}}, \bibinfo {author}
  {\bibfnamefont {R.}~\bibnamefont {Law}}, \bibinfo {author} {\bibfnamefont
  {D.}~\bibnamefont {Tilman}}, \bibinfo {author} {\bibfnamefont
  {M.}~\bibnamefont {Loreau}},\ and\ \bibinfo {author} {\bibfnamefont
  {A.}~\bibnamefont {Gonzalez}},\ }\bibfield  {title} {\bibinfo {title} {The
  metacommunity concept: A framework for multi-scale community ecology},\
  }\href {https://doi.org/10.1111/j.1461-0248.2004.00608.x} {\bibfield
  {journal} {\bibinfo  {journal} {Ecology Letters}\ }\textbf {\bibinfo {volume}
  {7}},\ \bibinfo {pages} {601} (\bibinfo {year} {2004})}\BibitemShut {NoStop}%
\bibitem [{\citenamefont {Gilarranz}\ and\ \citenamefont
  {Bascompte}(2012)}]{gilarranz2012spatial}%
  \BibitemOpen
  \bibfield  {author} {\bibinfo {author} {\bibfnamefont {L.~J.}\ \bibnamefont
  {Gilarranz}}\ and\ \bibinfo {author} {\bibfnamefont {J.}~\bibnamefont
  {Bascompte}},\ }\bibfield  {title} {\bibinfo {title} {Spatial network
  structure and metapopulation persistence},\ }\href
  {https://doi.org/10.1016/j.jtbi.2011.11.027} {\bibfield  {journal} {\bibinfo
  {journal} {Journal of Theoretical Biology}\ }\textbf {\bibinfo {volume}
  {297}},\ \bibinfo {pages} {11} (\bibinfo {year} {2012})}\BibitemShut
  {NoStop}%
\bibitem [{\citenamefont {Gravel}\ \emph {et~al.}(2016)\citenamefont {Gravel},
  \citenamefont {Massol},\ and\ \citenamefont {Leibold}}]{gravel2016stability}%
  \BibitemOpen
  \bibfield  {author} {\bibinfo {author} {\bibfnamefont {D.}~\bibnamefont
  {Gravel}}, \bibinfo {author} {\bibfnamefont {F.}~\bibnamefont {Massol}},\
  and\ \bibinfo {author} {\bibfnamefont {M.~A.}\ \bibnamefont {Leibold}},\
  }\bibfield  {title} {\bibinfo {title} {Stability and complexity in model
  meta-ecosystems},\ }\href {https://doi.org/10.1038/ncomms12457} {\bibfield
  {journal} {\bibinfo  {journal} {Nat Commun}\ }\textbf {\bibinfo {volume}
  {7}},\ \bibinfo {pages} {12457} (\bibinfo {year} {2016})}\BibitemShut
  {NoStop}%
\bibitem [{\citenamefont {Baron}\ and\ \citenamefont
  {Galla}(2020)}]{baron2020dispersalinduced}%
  \BibitemOpen
  \bibfield  {author} {\bibinfo {author} {\bibfnamefont {J.~W.}\ \bibnamefont
  {Baron}}\ and\ \bibinfo {author} {\bibfnamefont {T.}~\bibnamefont {Galla}},\
  }\bibfield  {title} {\bibinfo {title} {Dispersal-induced instability in
  complex ecosystems},\ }\href {https://doi.org/10.1038/s41467-020-19824-4}
  {\bibfield  {journal} {\bibinfo  {journal} {Nat Commun}\ }\textbf {\bibinfo
  {volume} {11}},\ \bibinfo {pages} {6032} (\bibinfo {year}
  {2020})}\BibitemShut {NoStop}%
\bibitem [{\citenamefont {Pettersson}\ and\ \citenamefont
  {Jacobi}(2021)}]{pettersson2021spatial}%
  \BibitemOpen
  \bibfield  {author} {\bibinfo {author} {\bibfnamefont {S.}~\bibnamefont
  {Pettersson}}\ and\ \bibinfo {author} {\bibfnamefont {M.~N.}\ \bibnamefont
  {Jacobi}},\ }\bibfield  {title} {\bibinfo {title} {Spatial heterogeneity
  enhance robustness of large multi-species ecosystems},\ }\href
  {https://doi.org/10.1371/journal.pcbi.1008899} {\bibfield  {journal}
  {\bibinfo  {journal} {PLOS Computational Biology}\ }\textbf {\bibinfo
  {volume} {17}},\ \bibinfo {pages} {e1008899} (\bibinfo {year}
  {2021})}\BibitemShut {NoStop}%
\bibitem [{\citenamefont {Garcia~Lorenzana}\ \emph {et~al.}(2024)\citenamefont
  {Garcia~Lorenzana}, \citenamefont {Altieri},\ and\ \citenamefont
  {Biroli}}]{garcialorenzana2024interactions}%
  \BibitemOpen
  \bibfield  {author} {\bibinfo {author} {\bibfnamefont {G.}~\bibnamefont
  {Garcia~Lorenzana}}, \bibinfo {author} {\bibfnamefont {A.}~\bibnamefont
  {Altieri}},\ and\ \bibinfo {author} {\bibfnamefont {G.}~\bibnamefont
  {Biroli}},\ }\bibfield  {title} {\bibinfo {title} {Interactions and
  {{Migration Rescuing Ecological Diversity}}},\ }\href
  {https://doi.org/10.1103/PRXLife.2.013014} {\bibfield  {journal} {\bibinfo
  {journal} {PRX Life}\ }\textbf {\bibinfo {volume} {2}},\ \bibinfo {pages}
  {013014} (\bibinfo {year} {2024})}\BibitemShut {NoStop}%
\bibitem [{\citenamefont {Hanski}\ and\ \citenamefont
  {Ovaskainen}(2000)}]{hanski2000metapopulation}%
  \BibitemOpen
  \bibfield  {author} {\bibinfo {author} {\bibfnamefont {I.}~\bibnamefont
  {Hanski}}\ and\ \bibinfo {author} {\bibfnamefont {O.}~\bibnamefont
  {Ovaskainen}},\ }\bibfield  {title} {\bibinfo {title} {The metapopulation
  capacity of a fragmented landscape},\ }\href
  {https://doi.org/10.1038/35008063} {\bibfield  {journal} {\bibinfo  {journal}
  {Nature}\ }\textbf {\bibinfo {volume} {404}},\ \bibinfo {pages} {755}
  (\bibinfo {year} {2000})}\BibitemShut {NoStop}%
\bibitem [{\citenamefont {Viswanathan}\ \emph {et~al.}(2008)\citenamefont
  {Viswanathan}, \citenamefont {Raposo},\ and\ \citenamefont
  {{\noopsort{luz}}{da Luz}}}]{viswanathan2008levy}%
  \BibitemOpen
  \bibfield  {author} {\bibinfo {author} {\bibfnamefont {G.~M.}\ \bibnamefont
  {Viswanathan}}, \bibinfo {author} {\bibfnamefont {E.~P.}\ \bibnamefont
  {Raposo}},\ and\ \bibinfo {author} {\bibfnamefont {M.~G.~E.}\ \bibnamefont
  {{\noopsort{luz}}{da Luz}}},\ }\bibfield  {title} {\bibinfo {title} {L{\'e}vy
  flights and superdiffusion in the context of biological encounters and random
  searches},\ }\href {https://doi.org/10.1016/j.plrev.2008.03.002} {\bibfield
  {journal} {\bibinfo  {journal} {Physics of Life Reviews}\ }\textbf {\bibinfo
  {volume} {5}},\ \bibinfo {pages} {133} (\bibinfo {year} {2008})}\BibitemShut
  {NoStop}%
\bibitem [{\citenamefont {Clobert}\ \emph {et~al.}(2012)\citenamefont
  {Clobert}, \citenamefont {Baguette}, \citenamefont {Benton},\ and\
  \citenamefont {Bullock}}]{clobert2012dispersal}%
  \BibitemOpen
  \bibfield  {author} {\bibinfo {author} {\bibfnamefont {J.}~\bibnamefont
  {Clobert}}, \bibinfo {author} {\bibfnamefont {M.}~\bibnamefont {Baguette}},
  \bibinfo {author} {\bibfnamefont {T.~G.}\ \bibnamefont {Benton}},\ and\
  \bibinfo {author} {\bibfnamefont {J.~M.}\ \bibnamefont {Bullock}},\
  }\href@noop {} {\emph {\bibinfo {title} {Dispersal {{Ecology}} and
  {{Evolution}}}}}\ (\bibinfo  {publisher} {OUP Oxford},\ \bibinfo {year}
  {2012})\BibitemShut {NoStop}%
\bibitem [{\citenamefont {Stone}(2018)}]{stone2018feasibility}%
  \BibitemOpen
  \bibfield  {author} {\bibinfo {author} {\bibfnamefont {L.}~\bibnamefont
  {Stone}},\ }\bibfield  {title} {\bibinfo {title} {The feasibility and
  stability of large complex biological networks: A random matrix approach},\
  }\href {https://doi.org/10.1038/s41598-018-26486-2} {\bibfield  {journal}
  {\bibinfo  {journal} {Sci Rep}\ }\textbf {\bibinfo {volume} {8}},\ \bibinfo
  {pages} {8246} (\bibinfo {year} {2018})}\BibitemShut {NoStop}%
\bibitem [{\citenamefont {Allesina}\ and\ \citenamefont
  {Grilli}(2020)}]{allesina2020models}%
  \BibitemOpen
  \bibfield  {author} {\bibinfo {author} {\bibfnamefont {S.}~\bibnamefont
  {Allesina}}\ and\ \bibinfo {author} {\bibfnamefont {J.}~\bibnamefont
  {Grilli}},\ }\bibfield  {title} {\bibinfo {title} {Models for large
  ecological communities---a random matrix approach},\ }in\ \href
  {https://doi.org/10.1093/oso/9780198824282.003.0006} {\emph {\bibinfo
  {booktitle} {Theoretical {{Ecology}}: Concepts and Applications}}},\ \bibinfo
  {editor} {edited by\ \bibinfo {editor} {\bibfnamefont {K.~S.}\ \bibnamefont
  {McCann}}\ and\ \bibinfo {editor} {\bibfnamefont {G.}~\bibnamefont
  {Gellner}}}\ (\bibinfo  {publisher} {Oxford University Press},\ \bibinfo
  {year} {2020})\ p.~\bibinfo {pages} {0}\BibitemShut {NoStop}%
\bibitem [{\citenamefont {Nicoletti}\ \emph {et~al.}(2023)\citenamefont
  {Nicoletti}, \citenamefont {Padmanabha}, \citenamefont {Azaele},
  \citenamefont {Suweis}, \citenamefont {Rinaldo},\ and\ \citenamefont
  {Maritan}}]{nicoletti2023emergent}%
  \BibitemOpen
  \bibfield  {author} {\bibinfo {author} {\bibfnamefont {G.}~\bibnamefont
  {Nicoletti}}, \bibinfo {author} {\bibfnamefont {P.}~\bibnamefont
  {Padmanabha}}, \bibinfo {author} {\bibfnamefont {S.}~\bibnamefont {Azaele}},
  \bibinfo {author} {\bibfnamefont {S.}~\bibnamefont {Suweis}}, \bibinfo
  {author} {\bibfnamefont {A.}~\bibnamefont {Rinaldo}},\ and\ \bibinfo {author}
  {\bibfnamefont {A.}~\bibnamefont {Maritan}},\ }\bibfield  {title} {\bibinfo
  {title} {Emergent encoding of dispersal network topologies in spatial
  metapopulation models},\ }\href {https://doi.org/10.1073/pnas.2311548120}
  {\bibfield  {journal} {\bibinfo  {journal} {Proceedings of the National
  Academy of Sciences}\ }\textbf {\bibinfo {volume} {120}},\ \bibinfo {pages}
  {e2311548120} (\bibinfo {year} {2023})}\BibitemShut {NoStop}%
\bibitem [{\citenamefont {De~Domenico}\ \emph {et~al.}(2013)\citenamefont
  {De~Domenico}, \citenamefont {{Sol{\'e}-Ribalta}}, \citenamefont {Cozzo},
  \citenamefont {Kivel{\"a}}, \citenamefont {Moreno}, \citenamefont {Porter},
  \citenamefont {G{\'o}mez},\ and\ \citenamefont
  {Arenas}}]{dedomenico2013mathematical}%
  \BibitemOpen
  \bibfield  {author} {\bibinfo {author} {\bibfnamefont {M.}~\bibnamefont
  {De~Domenico}}, \bibinfo {author} {\bibfnamefont {A.}~\bibnamefont
  {{Sol{\'e}-Ribalta}}}, \bibinfo {author} {\bibfnamefont {E.}~\bibnamefont
  {Cozzo}}, \bibinfo {author} {\bibfnamefont {M.}~\bibnamefont {Kivel{\"a}}},
  \bibinfo {author} {\bibfnamefont {Y.}~\bibnamefont {Moreno}}, \bibinfo
  {author} {\bibfnamefont {M.~A.}\ \bibnamefont {Porter}}, \bibinfo {author}
  {\bibfnamefont {S.}~\bibnamefont {G{\'o}mez}},\ and\ \bibinfo {author}
  {\bibfnamefont {A.}~\bibnamefont {Arenas}},\ }\bibfield  {title} {\bibinfo
  {title} {Mathematical {{Formulation}} of {{Multilayer Networks}}},\ }\href
  {https://doi.org/10.1103/PhysRevX.3.041022} {\bibfield  {journal} {\bibinfo
  {journal} {Phys. Rev. X}\ }\textbf {\bibinfo {volume} {3}},\ \bibinfo {pages}
  {041022} (\bibinfo {year} {2013})}\BibitemShut {NoStop}%
\bibitem [{\citenamefont {Pilosof}\ \emph {et~al.}(2017)\citenamefont
  {Pilosof}, \citenamefont {Porter}, \citenamefont {Pascual},\ and\
  \citenamefont {K{\'e}fi}}]{pilosof2017multilayer}%
  \BibitemOpen
  \bibfield  {author} {\bibinfo {author} {\bibfnamefont {S.}~\bibnamefont
  {Pilosof}}, \bibinfo {author} {\bibfnamefont {M.~A.}\ \bibnamefont {Porter}},
  \bibinfo {author} {\bibfnamefont {M.}~\bibnamefont {Pascual}},\ and\ \bibinfo
  {author} {\bibfnamefont {S.}~\bibnamefont {K{\'e}fi}},\ }\bibfield  {title}
  {\bibinfo {title} {The multilayer nature of ecological networks},\ }\href
  {https://doi.org/10.1038/s41559-017-0101} {\bibfield  {journal} {\bibinfo
  {journal} {Nat Ecol Evol}\ }\textbf {\bibinfo {volume} {1}},\ \bibinfo
  {pages} {1} (\bibinfo {year} {2017})}\BibitemShut {NoStop}%
\bibitem [{\citenamefont {Brechtel}\ \emph {et~al.}(2018)\citenamefont
  {Brechtel}, \citenamefont {Gramlich}, \citenamefont {Ritterskamp},
  \citenamefont {Drossel},\ and\ \citenamefont {Gross}}]{brechtel2018master}%
  \BibitemOpen
  \bibfield  {author} {\bibinfo {author} {\bibfnamefont {A.}~\bibnamefont
  {Brechtel}}, \bibinfo {author} {\bibfnamefont {P.}~\bibnamefont {Gramlich}},
  \bibinfo {author} {\bibfnamefont {D.}~\bibnamefont {Ritterskamp}}, \bibinfo
  {author} {\bibfnamefont {B.}~\bibnamefont {Drossel}},\ and\ \bibinfo {author}
  {\bibfnamefont {T.}~\bibnamefont {Gross}},\ }\bibfield  {title} {\bibinfo
  {title} {Master stability functions reveal diffusion-driven pattern formation
  in networks},\ }\href {https://doi.org/10.1103/PhysRevE.97.032307} {\bibfield
   {journal} {\bibinfo  {journal} {Phys. Rev. E}\ }\textbf {\bibinfo {volume}
  {97}},\ \bibinfo {pages} {032307} (\bibinfo {year} {2018})}\BibitemShut
  {NoStop}%
\bibitem [{\citenamefont {Tao}\ \emph {et~al.}(2010)\citenamefont {Tao},
  \citenamefont {Vu},\ and\ \citenamefont {Krishnapur}}]{tao2010random}%
  \BibitemOpen
  \bibfield  {author} {\bibinfo {author} {\bibfnamefont {T.}~\bibnamefont
  {Tao}}, \bibinfo {author} {\bibfnamefont {V.}~\bibnamefont {Vu}},\ and\
  \bibinfo {author} {\bibfnamefont {M.}~\bibnamefont {Krishnapur}},\ }\bibfield
   {title} {\bibinfo {title} {Random matrices: {{Universality}} of {{ESDs}} and
  the circular law},\ }\href {https://doi.org/10.1214/10-AOP534} {\bibfield
  {journal} {\bibinfo  {journal} {The Annals of Probability}\ }\textbf
  {\bibinfo {volume} {38}},\ \bibinfo {pages} {2023} (\bibinfo {year}
  {2010})}\BibitemShut {NoStop}%
\bibitem [{\citenamefont {Barab{\'a}s}\ \emph {et~al.}(2017)\citenamefont
  {Barab{\'a}s}, \citenamefont {{Michalska-Smith}},\ and\ \citenamefont
  {Allesina}}]{barabas2017selfregulation}%
  \BibitemOpen
  \bibfield  {author} {\bibinfo {author} {\bibfnamefont {G.}~\bibnamefont
  {Barab{\'a}s}}, \bibinfo {author} {\bibfnamefont {M.~J.}\ \bibnamefont
  {{Michalska-Smith}}},\ and\ \bibinfo {author} {\bibfnamefont
  {S.}~\bibnamefont {Allesina}},\ }\bibfield  {title} {\bibinfo {title}
  {Self-regulation and the stability of large ecological networks},\ }\href
  {https://doi.org/10.1038/s41559-017-0357-6} {\bibfield  {journal} {\bibinfo
  {journal} {Nat Ecol Evol}\ }\textbf {\bibinfo {volume} {1}},\ \bibinfo
  {pages} {1870} (\bibinfo {year} {2017})}\BibitemShut {NoStop}%
\bibitem [{\citenamefont {Niebuhr}\ \emph {et~al.}(2015)\citenamefont
  {Niebuhr}, \citenamefont {Wosniack}, \citenamefont {Santos}, \citenamefont
  {Raposo}, \citenamefont {Viswanathan}, \citenamefont {{\noopsort{luz}}{da
  Luz}},\ and\ \citenamefont {Pie}}]{niebuhr2015survival}%
  \BibitemOpen
  \bibfield  {author} {\bibinfo {author} {\bibfnamefont {B.~B.~S.}\
  \bibnamefont {Niebuhr}}, \bibinfo {author} {\bibfnamefont {M.~E.}\
  \bibnamefont {Wosniack}}, \bibinfo {author} {\bibfnamefont {M.~C.}\
  \bibnamefont {Santos}}, \bibinfo {author} {\bibfnamefont {E.~P.}\
  \bibnamefont {Raposo}}, \bibinfo {author} {\bibfnamefont {G.~M.}\
  \bibnamefont {Viswanathan}}, \bibinfo {author} {\bibfnamefont {M.~G.~E.}\
  \bibnamefont {{\noopsort{luz}}{da Luz}}},\ and\ \bibinfo {author}
  {\bibfnamefont {M.~R.}\ \bibnamefont {Pie}},\ }\bibfield  {title} {\bibinfo
  {title} {Survival in patchy landscapes: The interplay between dispersal,
  habitat loss and fragmentation},\ }\href {https://doi.org/10.1038/srep11898}
  {\bibfield  {journal} {\bibinfo  {journal} {Sci Rep}\ }\textbf {\bibinfo
  {volume} {5}},\ \bibinfo {pages} {11898} (\bibinfo {year}
  {2015})}\BibitemShut {NoStop}%
\bibitem [{\citenamefont {Bertassello}\ \emph {et~al.}(2020)\citenamefont
  {Bertassello}, \citenamefont {Aubeneau}, \citenamefont {Botter},
  \citenamefont {Jawitz},\ and\ \citenamefont {Rao}}]{bertassello2020emergent}%
  \BibitemOpen
  \bibfield  {author} {\bibinfo {author} {\bibfnamefont {L.~E.}\ \bibnamefont
  {Bertassello}}, \bibinfo {author} {\bibfnamefont {A.~F.}\ \bibnamefont
  {Aubeneau}}, \bibinfo {author} {\bibfnamefont {G.}~\bibnamefont {Botter}},
  \bibinfo {author} {\bibfnamefont {J.~W.}\ \bibnamefont {Jawitz}},\ and\
  \bibinfo {author} {\bibfnamefont {P.~S.~C.}\ \bibnamefont {Rao}},\ }\bibfield
   {title} {\bibinfo {title} {Emergent dispersal networks in dynamic
  wetlandscapes},\ }\href {https://doi.org/10.1038/s41598-020-71739-8}
  {\bibfield  {journal} {\bibinfo  {journal} {Sci Rep}\ }\textbf {\bibinfo
  {volume} {10}},\ \bibinfo {pages} {14696} (\bibinfo {year}
  {2020})}\BibitemShut {NoStop}%
\bibitem [{\citenamefont {Padmanabha}\ \emph {et~al.}(2024)\citenamefont
  {Padmanabha}, \citenamefont {Nicoletti}, \citenamefont {Bernardi},
  \citenamefont {Suweis}, \citenamefont {Azaele}, \citenamefont {Rinaldo},\
  and\ \citenamefont {Maritan}}]{padmanabha2024spatially}%
  \BibitemOpen
  \bibfield  {author} {\bibinfo {author} {\bibfnamefont {P.}~\bibnamefont
  {Padmanabha}}, \bibinfo {author} {\bibfnamefont {G.}~\bibnamefont
  {Nicoletti}}, \bibinfo {author} {\bibfnamefont {D.}~\bibnamefont {Bernardi}},
  \bibinfo {author} {\bibfnamefont {S.}~\bibnamefont {Suweis}}, \bibinfo
  {author} {\bibfnamefont {S.}~\bibnamefont {Azaele}}, \bibinfo {author}
  {\bibfnamefont {A.}~\bibnamefont {Rinaldo}},\ and\ \bibinfo {author}
  {\bibfnamefont {A.}~\bibnamefont {Maritan}},\ }\href
  {https://doi.org/10.48550/arXiv.2404.09908} {\bibinfo {title} {Spatially
  disordered environments stabilize competitive metacommunities}} (\bibinfo
  {year} {2024}),\ \Eprint {https://arxiv.org/abs/2404.09908} {arXiv:2404.09908
  [cond-mat, q-bio]} \BibitemShut {NoStop}%
\bibitem [{\citenamefont {Gilarranz}(2020)}]{gilarranz2020generic}%
  \BibitemOpen
  \bibfield  {author} {\bibinfo {author} {\bibfnamefont {L.~J.}\ \bibnamefont
  {Gilarranz}},\ }\bibfield  {title} {\bibinfo {title} {Generic {{Emergence}}
  of {{Modularity}} in {{Spatial Networks}}},\ }\href
  {https://doi.org/10.1038/s41598-020-65669-8} {\bibfield  {journal} {\bibinfo
  {journal} {Sci Rep}\ }\textbf {\bibinfo {volume} {10}},\ \bibinfo {pages}
  {8708} (\bibinfo {year} {2020})}\BibitemShut {NoStop}%
\bibitem [{\citenamefont {Grilli}\ \emph {et~al.}(2015)\citenamefont {Grilli},
  \citenamefont {Barab{\'a}s},\ and\ \citenamefont
  {Allesina}}]{grilli2015metapopulation}%
  \BibitemOpen
  \bibfield  {author} {\bibinfo {author} {\bibfnamefont {J.}~\bibnamefont
  {Grilli}}, \bibinfo {author} {\bibfnamefont {G.}~\bibnamefont
  {Barab{\'a}s}},\ and\ \bibinfo {author} {\bibfnamefont {S.}~\bibnamefont
  {Allesina}},\ }\bibfield  {title} {\bibinfo {title} {Metapopulation
  {{Persistence}} in {{Random Fragmented Landscapes}}},\ }\href
  {https://doi.org/10.1371/journal.pcbi.1004251} {\bibfield  {journal}
  {\bibinfo  {journal} {PLOS Computational Biology}\ }\textbf {\bibinfo
  {volume} {11}},\ \bibinfo {pages} {e1004251} (\bibinfo {year}
  {2015})}\BibitemShut {NoStop}%
\bibitem [{\citenamefont {Newman}(2009)}]{newman2009random}%
  \BibitemOpen
  \bibfield  {author} {\bibinfo {author} {\bibfnamefont {M.~E.~J.}\
  \bibnamefont {Newman}},\ }\bibfield  {title} {\bibinfo {title} {Random
  {{Graphs}} with {{Clustering}}},\ }\href
  {https://doi.org/10.1103/PhysRevLett.103.058701} {\bibfield  {journal}
  {\bibinfo  {journal} {Phys. Rev. Lett.}\ }\textbf {\bibinfo {volume} {103}},\
  \bibinfo {pages} {058701} (\bibinfo {year} {2009})}\BibitemShut {NoStop}%
\bibitem [{\citenamefont {Newman}(2018)}]{newman2018networks}%
  \BibitemOpen
  \bibfield  {author} {\bibinfo {author} {\bibfnamefont {M.}~\bibnamefont
  {Newman}},\ }\href@noop {} {\emph {\bibinfo {title} {Networks}}}\ (\bibinfo
  {publisher} {Oxford University Press},\ \bibinfo {year} {2018})\BibitemShut
  {NoStop}%
\bibitem [{\citenamefont {Tishby}\ \emph
  {et~al.}(2018{\natexlab{a}})\citenamefont {Tishby}, \citenamefont {Biham},
  \citenamefont {Katzav},\ and\ \citenamefont
  {K{\"u}hn}}]{tishby2018revealing}%
  \BibitemOpen
  \bibfield  {author} {\bibinfo {author} {\bibfnamefont {I.}~\bibnamefont
  {Tishby}}, \bibinfo {author} {\bibfnamefont {O.}~\bibnamefont {Biham}},
  \bibinfo {author} {\bibfnamefont {E.}~\bibnamefont {Katzav}},\ and\ \bibinfo
  {author} {\bibfnamefont {R.}~\bibnamefont {K{\"u}hn}},\ }\bibfield  {title}
  {\bibinfo {title} {Revealing the microstructure of the giant component in
  random graph ensembles},\ }\href {https://doi.org/10.1103/PhysRevE.97.042318}
  {\bibfield  {journal} {\bibinfo  {journal} {Phys. Rev. E}\ }\textbf {\bibinfo
  {volume} {97}},\ \bibinfo {pages} {042318} (\bibinfo {year}
  {2018}{\natexlab{a}})}\BibitemShut {NoStop}%
\bibitem [{\citenamefont {Tishby}\ \emph
  {et~al.}(2018{\natexlab{b}})\citenamefont {Tishby}, \citenamefont {Biham},
  \citenamefont {K{\"u}hn},\ and\ \citenamefont
  {Katzav}}]{tishby2018statistical}%
  \BibitemOpen
  \bibfield  {author} {\bibinfo {author} {\bibfnamefont {I.}~\bibnamefont
  {Tishby}}, \bibinfo {author} {\bibfnamefont {O.}~\bibnamefont {Biham}},
  \bibinfo {author} {\bibfnamefont {R.}~\bibnamefont {K{\"u}hn}},\ and\
  \bibinfo {author} {\bibfnamefont {E.}~\bibnamefont {Katzav}},\ }\bibfield
  {title} {\bibinfo {title} {Statistical analysis of articulation points in
  configuration model networks},\ }\href
  {https://doi.org/10.1103/PhysRevE.98.062301} {\bibfield  {journal} {\bibinfo
  {journal} {Phys. Rev. E}\ }\textbf {\bibinfo {volume} {98}},\ \bibinfo
  {pages} {062301} (\bibinfo {year} {2018}{\natexlab{b}})}\BibitemShut
  {NoStop}%
\bibitem [{\citenamefont {Tishby}\ \emph {et~al.}(2019)\citenamefont {Tishby},
  \citenamefont {Biham}, \citenamefont {Katzav},\ and\ \citenamefont
  {K{\"u}hn}}]{tishby2019generating}%
  \BibitemOpen
  \bibfield  {author} {\bibinfo {author} {\bibfnamefont {I.}~\bibnamefont
  {Tishby}}, \bibinfo {author} {\bibfnamefont {O.}~\bibnamefont {Biham}},
  \bibinfo {author} {\bibfnamefont {E.}~\bibnamefont {Katzav}},\ and\ \bibinfo
  {author} {\bibfnamefont {R.}~\bibnamefont {K{\"u}hn}},\ }\bibfield  {title}
  {\bibinfo {title} {Generating random networks that consist of a single
  connected component with a given degree distribution},\ }\href
  {https://doi.org/10.1103/PhysRevE.99.042308} {\bibfield  {journal} {\bibinfo
  {journal} {Phys. Rev. E}\ }\textbf {\bibinfo {volume} {99}},\ \bibinfo
  {pages} {042308} (\bibinfo {year} {2019})}\BibitemShut {NoStop}%
\bibitem [{\citenamefont {Ghavasieh}\ and\ \citenamefont
  {De~Domenico}(2024)}]{ghavasieh2024diversity}%
  \BibitemOpen
  \bibfield  {author} {\bibinfo {author} {\bibfnamefont {A.}~\bibnamefont
  {Ghavasieh}}\ and\ \bibinfo {author} {\bibfnamefont {M.}~\bibnamefont
  {De~Domenico}},\ }\bibfield  {title} {\bibinfo {title} {Diversity of
  information pathways drives sparsity in real-world networks},\ }\href
  {https://doi.org/10.1038/s41567-023-02330-x} {\bibfield  {journal} {\bibinfo
  {journal} {Nat. Phys.}\ ,\ \bibinfo {pages} {1}} (\bibinfo {year}
  {2024})}\BibitemShut {NoStop}%
\bibitem [{\citenamefont {Venturelli}\ \emph {et~al.}(2018)\citenamefont
  {Venturelli}, \citenamefont {Carr}, \citenamefont {Fisher}, \citenamefont
  {Hsu}, \citenamefont {Lau}, \citenamefont {Bowen}, \citenamefont {Hromada},
  \citenamefont {Northen},\ and\ \citenamefont
  {Arkin}}]{venturelli2018deciphering}%
  \BibitemOpen
  \bibfield  {author} {\bibinfo {author} {\bibfnamefont {O.~S.}\ \bibnamefont
  {Venturelli}}, \bibinfo {author} {\bibfnamefont {A.~V.}\ \bibnamefont
  {Carr}}, \bibinfo {author} {\bibfnamefont {G.}~\bibnamefont {Fisher}},
  \bibinfo {author} {\bibfnamefont {R.~H.}\ \bibnamefont {Hsu}}, \bibinfo
  {author} {\bibfnamefont {R.}~\bibnamefont {Lau}}, \bibinfo {author}
  {\bibfnamefont {B.~P.}\ \bibnamefont {Bowen}}, \bibinfo {author}
  {\bibfnamefont {S.}~\bibnamefont {Hromada}}, \bibinfo {author} {\bibfnamefont
  {T.}~\bibnamefont {Northen}},\ and\ \bibinfo {author} {\bibfnamefont {A.~P.}\
  \bibnamefont {Arkin}},\ }\bibfield  {title} {\bibinfo {title} {Deciphering
  microbial interactions in synthetic human gut microbiome communities},\
  }\href {https://doi.org/10.15252/msb.20178157} {\bibfield  {journal}
  {\bibinfo  {journal} {Molecular Systems Biology}\ }\textbf {\bibinfo {volume}
  {14}},\ \bibinfo {pages} {e8157} (\bibinfo {year} {2018})}\BibitemShut
  {NoStop}%
\bibitem [{\citenamefont {Kurkjian}(2019)}]{kurkjian2019metapopulation}%
  \BibitemOpen
  \bibfield  {author} {\bibinfo {author} {\bibfnamefont {H.~M.}\ \bibnamefont
  {Kurkjian}},\ }\bibfield  {title} {\bibinfo {title} {The {{Metapopulation
  Microcosm Plate}}: {{A}} modified 96-well plate for use in microbial
  metapopulation experiments},\ }\href
  {https://doi.org/10.1111/2041-210X.13116} {\bibfield  {journal} {\bibinfo
  {journal} {Methods in Ecology and Evolution}\ }\textbf {\bibinfo {volume}
  {10}},\ \bibinfo {pages} {162} (\bibinfo {year} {2019})}\BibitemShut
  {NoStop}%
\bibitem [{\citenamefont {Larsen}\ and\ \citenamefont
  {Hargreaves}(2020)}]{larsen2020miniaturizing}%
  \BibitemOpen
  \bibfield  {author} {\bibinfo {author} {\bibfnamefont {C.~D.}\ \bibnamefont
  {Larsen}}\ and\ \bibinfo {author} {\bibfnamefont {A.~L.}\ \bibnamefont
  {Hargreaves}},\ }\bibfield  {title} {\bibinfo {title} {Miniaturizing
  landscapes to understand species distributions},\ }\href
  {https://doi.org/10.1111/ecog.04959} {\bibfield  {journal} {\bibinfo
  {journal} {Ecography}\ }\textbf {\bibinfo {volume} {43}},\ \bibinfo {pages}
  {1625} (\bibinfo {year} {2020})}\BibitemShut {NoStop}%
\bibitem [{\citenamefont {Watts}\ and\ \citenamefont
  {Strogatz}(1998)}]{watts1998collective}%
  \BibitemOpen
  \bibfield  {author} {\bibinfo {author} {\bibfnamefont {D.~J.}\ \bibnamefont
  {Watts}}\ and\ \bibinfo {author} {\bibfnamefont {S.~H.}\ \bibnamefont
  {Strogatz}},\ }\bibfield  {title} {\bibinfo {title} {Collective dynamics of
  `small-world' networks},\ }\href {https://doi.org/10.1038/30918} {\bibfield
  {journal} {\bibinfo  {journal} {Nature}\ }\textbf {\bibinfo {volume} {393}},\
  \bibinfo {pages} {440} (\bibinfo {year} {1998})}\BibitemShut {NoStop}%
\bibitem [{\citenamefont {Mouillot}(2007)}]{mouillot2007nicheassembly}%
  \BibitemOpen
  \bibfield  {author} {\bibinfo {author} {\bibfnamefont {D.}~\bibnamefont
  {Mouillot}},\ }\bibfield  {title} {\bibinfo {title} {Niche-{{Assembly}} vs.
  {{Dispersal-Assembly Rules}} in {{Coastal Fish Metacommunities}}:
  {{Implications}} for {{Management}} of {{Biodiversity}} in {{Brackish
  Lagoons}}},\ }\href@noop {} {\bibfield  {journal} {\bibinfo  {journal}
  {Journal of Applied Ecology}\ }\textbf {\bibinfo {volume} {44}},\ \bibinfo
  {pages} {760} (\bibinfo {year} {2007})},\ \Eprint
  {https://arxiv.org/abs/4539295} {4539295} \BibitemShut {NoStop}%
\bibitem [{\citenamefont {Gravel}\ \emph {et~al.}(2010)\citenamefont {Gravel},
  \citenamefont {Guichard}, \citenamefont {Loreau},\ and\ \citenamefont
  {Mouquet}}]{gravel2010source}%
  \BibitemOpen
  \bibfield  {author} {\bibinfo {author} {\bibfnamefont {D.}~\bibnamefont
  {Gravel}}, \bibinfo {author} {\bibfnamefont {F.}~\bibnamefont {Guichard}},
  \bibinfo {author} {\bibfnamefont {M.}~\bibnamefont {Loreau}},\ and\ \bibinfo
  {author} {\bibfnamefont {N.}~\bibnamefont {Mouquet}},\ }\bibfield  {title}
  {\bibinfo {title} {Source and sink dynamics in meta-ecosystems},\ }\href
  {https://doi.org/10.1890/09-0843.1} {\bibfield  {journal} {\bibinfo
  {journal} {Ecology}\ }\textbf {\bibinfo {volume} {91}},\ \bibinfo {pages}
  {2172} (\bibinfo {year} {2010})}\BibitemShut {NoStop}%
\bibitem [{\citenamefont {Loke}\ and\ \citenamefont
  {Chisholm}(2023)}]{loke2023unveiling}%
  \BibitemOpen
  \bibfield  {author} {\bibinfo {author} {\bibfnamefont {L.~H.~L.}\
  \bibnamefont {Loke}}\ and\ \bibinfo {author} {\bibfnamefont {R.~A.}\
  \bibnamefont {Chisholm}},\ }\bibfield  {title} {\bibinfo {title} {Unveiling
  the transition from niche to dispersal assembly in ecology},\ }\href
  {https://doi.org/10.1038/s41586-023-06161-x} {\bibfield  {journal} {\bibinfo
  {journal} {Nature}\ }\textbf {\bibinfo {volume} {618}},\ \bibinfo {pages}
  {537} (\bibinfo {year} {2023})}\BibitemShut {NoStop}%
\bibitem [{\citenamefont {Hanski}(1998)}]{hanski1998metapopulation}%
  \BibitemOpen
  \bibfield  {author} {\bibinfo {author} {\bibfnamefont {I.}~\bibnamefont
  {Hanski}},\ }\bibfield  {title} {\bibinfo {title} {Metapopulation dynamics},\
  }\href {https://doi.org/10.1038/23876} {\bibfield  {journal} {\bibinfo
  {journal} {Nature}\ }\textbf {\bibinfo {volume} {396}},\ \bibinfo {pages}
  {41} (\bibinfo {year} {1998})}\BibitemShut {NoStop}%
\bibitem [{\citenamefont {Wennekes}\ \emph {et~al.}(2012)\citenamefont
  {Wennekes}, \citenamefont {Rosindell},\ and\ \citenamefont
  {Etienne}}]{wennekes2012neutral}%
  \BibitemOpen
  \bibfield  {author} {\bibinfo {author} {\bibfnamefont {P.~L.}\ \bibnamefont
  {Wennekes}}, \bibinfo {author} {\bibfnamefont {J.}~\bibnamefont
  {Rosindell}},\ and\ \bibinfo {author} {\bibfnamefont {R.~S.}\ \bibnamefont
  {Etienne}},\ }\bibfield  {title} {\bibinfo {title} {The {{Neutral}}---{{Niche
  Debate}}: {{A Philosophical Perspective}}},\ }\href
  {https://doi.org/10.1007/s10441-012-9144-6} {\bibfield  {journal} {\bibinfo
  {journal} {Acta Biotheor}\ }\textbf {\bibinfo {volume} {60}},\ \bibinfo
  {pages} {257} (\bibinfo {year} {2012})}\BibitemShut {NoStop}%
\bibitem [{\citenamefont {Reppas}\ \emph {et~al.}(2015)\citenamefont {Reppas},
  \citenamefont {Spiliotis},\ and\ \citenamefont {Siettos}}]{reppas2015tuning}%
  \BibitemOpen
  \bibfield  {author} {\bibinfo {author} {\bibfnamefont {A.~I.}\ \bibnamefont
  {Reppas}}, \bibinfo {author} {\bibfnamefont {K.}~\bibnamefont {Spiliotis}},\
  and\ \bibinfo {author} {\bibfnamefont {C.~I.}\ \bibnamefont {Siettos}},\
  }\bibfield  {title} {\bibinfo {title} {Tuning the average path length of
  complex networks and its influence to the emergent dynamics of the
  majority-rule model},\ }\href {https://doi.org/10.1016/j.matcom.2014.09.005}
  {\bibfield  {journal} {\bibinfo  {journal} {Mathematics and Computers in
  Simulation}\ }\textbf {\bibinfo {volume} {109}},\ \bibinfo {pages} {186}
  (\bibinfo {year} {2015})}\BibitemShut {NoStop}%
\bibitem [{\citenamefont {Dall}\ and\ \citenamefont
  {Christensen}(2002)}]{dall2002random}%
  \BibitemOpen
  \bibfield  {author} {\bibinfo {author} {\bibfnamefont {J.}~\bibnamefont
  {Dall}}\ and\ \bibinfo {author} {\bibfnamefont {M.}~\bibnamefont
  {Christensen}},\ }\bibfield  {title} {\bibinfo {title} {Random geometric
  graphs},\ }\href {https://doi.org/10.1103/PhysRevE.66.016121} {\bibfield
  {journal} {\bibinfo  {journal} {Phys. Rev. E}\ }\textbf {\bibinfo {volume}
  {66}},\ \bibinfo {pages} {016121} (\bibinfo {year} {2002})}\BibitemShut
  {NoStop}%
\bibitem [{\citenamefont {Herrmann}\ \emph {et~al.}(2003)\citenamefont
  {Herrmann}, \citenamefont {Barth{\'e}lemy},\ and\ \citenamefont
  {Provero}}]{herrmann2003connectivity}%
  \BibitemOpen
  \bibfield  {author} {\bibinfo {author} {\bibfnamefont {C.}~\bibnamefont
  {Herrmann}}, \bibinfo {author} {\bibfnamefont {M.}~\bibnamefont
  {Barth{\'e}lemy}},\ and\ \bibinfo {author} {\bibfnamefont {P.}~\bibnamefont
  {Provero}},\ }\bibfield  {title} {\bibinfo {title} {Connectivity distribution
  of spatial networks},\ }\href {https://doi.org/10.1103/PhysRevE.68.026128}
  {\bibfield  {journal} {\bibinfo  {journal} {Phys. Rev. E}\ }\textbf {\bibinfo
  {volume} {68}},\ \bibinfo {pages} {026128} (\bibinfo {year}
  {2003})}\BibitemShut {NoStop}%
\bibitem [{\citenamefont {Plaszczynski}\ \emph {et~al.}(2022)\citenamefont
  {Plaszczynski}, \citenamefont {Nakamura}, \citenamefont {Deroulers},
  \citenamefont {Grammaticos},\ and\ \citenamefont
  {Badoual}}]{plaszczynski2022levy}%
  \BibitemOpen
  \bibfield  {author} {\bibinfo {author} {\bibfnamefont {S.}~\bibnamefont
  {Plaszczynski}}, \bibinfo {author} {\bibfnamefont {G.}~\bibnamefont
  {Nakamura}}, \bibinfo {author} {\bibfnamefont {C.}~\bibnamefont {Deroulers}},
  \bibinfo {author} {\bibfnamefont {B.}~\bibnamefont {Grammaticos}},\ and\
  \bibinfo {author} {\bibfnamefont {M.}~\bibnamefont {Badoual}},\ }\bibfield
  {title} {\bibinfo {title} {Levy geometric graphs},\ }\href
  {https://doi.org/10.1103/PhysRevE.105.054151} {\bibfield  {journal} {\bibinfo
   {journal} {Phys. Rev. E}\ }\textbf {\bibinfo {volume} {105}},\ \bibinfo
  {pages} {054151} (\bibinfo {year} {2022})}\BibitemShut {NoStop}%
\bibitem [{\citenamefont {Althagafi}\ and\ \citenamefont
  {Petrovskii}(2021)}]{althagafi2021metapopulation}%
  \BibitemOpen
  \bibfield  {author} {\bibinfo {author} {\bibfnamefont {H.}~\bibnamefont
  {Althagafi}}\ and\ \bibinfo {author} {\bibfnamefont {S.}~\bibnamefont
  {Petrovskii}},\ }\bibfield  {title} {\bibinfo {title} {Metapopulation
  {{Persistence}} and {{Extinction}} in a {{Fragmented Random Habitat}}: {{A
  Simulation Study}}},\ }\href {https://doi.org/10.3390/math9182202} {\bibfield
   {journal} {\bibinfo  {journal} {Mathematics}\ }\textbf {\bibinfo {volume}
  {9}},\ \bibinfo {pages} {2202} (\bibinfo {year} {2021})}\BibitemShut
  {NoStop}%
\bibitem [{\citenamefont {Tilman}\ \emph {et~al.}(1994)\citenamefont {Tilman},
  \citenamefont {{May, Robert M.}}, \citenamefont {{Lehman, Clarence L.}},\
  and\ \citenamefont {{Nowak, Martin A.}}}]{tilman1994habitat}%
  \BibitemOpen
  \bibfield  {author} {\bibinfo {author} {\bibfnamefont {D.}~\bibnamefont
  {Tilman}}, \bibinfo {author} {\bibnamefont {{May, Robert M.}}}, \bibinfo
  {author} {\bibnamefont {{Lehman, Clarence L.}}},\ and\ \bibinfo {author}
  {\bibnamefont {{Nowak, Martin A.}}},\ }\bibfield  {title} {\bibinfo {title}
  {Habitat destruction and the extinction debt},\ }\href
  {https://doi.org/10.1038/371065a0} {\bibfield  {journal} {\bibinfo  {journal}
  {Nature}\ }\textbf {\bibinfo {volume} {371}},\ \bibinfo {pages} {65}
  (\bibinfo {year} {1994})}\BibitemShut {NoStop}%
\bibitem [{\citenamefont {Brooks}\ \emph {et~al.}(2002)\citenamefont {Brooks},
  \citenamefont {Mittermeier}, \citenamefont {Mittermeier}, \citenamefont
  {Da~Fonseca}, \citenamefont {Rylands}, \citenamefont {Konstant},
  \citenamefont {Flick}, \citenamefont {Pilgrim}, \citenamefont {Oldfield},
  \citenamefont {Magin},\ and\ \citenamefont
  {{Hilton-Taylor}}}]{brooks2002habitat}%
  \BibitemOpen
  \bibfield  {author} {\bibinfo {author} {\bibfnamefont {T.~M.}\ \bibnamefont
  {Brooks}}, \bibinfo {author} {\bibfnamefont {R.~A.}\ \bibnamefont
  {Mittermeier}}, \bibinfo {author} {\bibfnamefont {C.~G.}\ \bibnamefont
  {Mittermeier}}, \bibinfo {author} {\bibfnamefont {G.~A.~B.}\ \bibnamefont
  {Da~Fonseca}}, \bibinfo {author} {\bibfnamefont {A.~B.}\ \bibnamefont
  {Rylands}}, \bibinfo {author} {\bibfnamefont {W.~R.}\ \bibnamefont
  {Konstant}}, \bibinfo {author} {\bibfnamefont {P.}~\bibnamefont {Flick}},
  \bibinfo {author} {\bibfnamefont {J.}~\bibnamefont {Pilgrim}}, \bibinfo
  {author} {\bibfnamefont {S.}~\bibnamefont {Oldfield}}, \bibinfo {author}
  {\bibfnamefont {G.}~\bibnamefont {Magin}},\ and\ \bibinfo {author}
  {\bibfnamefont {C.}~\bibnamefont {{Hilton-Taylor}}},\ }\bibfield  {title}
  {\bibinfo {title} {Habitat {{Loss}} and {{Extinction}} in the {{Hotspots}} of
  {{Biodiversity}}},\ }\href {https://doi.org/10.1046/j.1523-1739.2002.00530.x}
  {\bibfield  {journal} {\bibinfo  {journal} {Conservation Biology}\ }\textbf
  {\bibinfo {volume} {16}},\ \bibinfo {pages} {909} (\bibinfo {year}
  {2002})}\BibitemShut {NoStop}%
\bibitem [{\citenamefont {Hatton}\ \emph {et~al.}(2024)\citenamefont {Hatton},
  \citenamefont {Mazzarisi}, \citenamefont {Altieri},\ and\ \citenamefont
  {Smerlak}}]{hatton2024diversity}%
  \BibitemOpen
  \bibfield  {author} {\bibinfo {author} {\bibfnamefont {I.~A.}\ \bibnamefont
  {Hatton}}, \bibinfo {author} {\bibfnamefont {O.}~\bibnamefont {Mazzarisi}},
  \bibinfo {author} {\bibfnamefont {A.}~\bibnamefont {Altieri}},\ and\ \bibinfo
  {author} {\bibfnamefont {M.}~\bibnamefont {Smerlak}},\ }\bibfield  {title}
  {\bibinfo {title} {Diversity begets stability: {{Sublinear}} growth and
  competitive coexistence across ecosystems},\ }\href
  {https://doi.org/10.1126/science.adg8488} {\bibfield  {journal} {\bibinfo
  {journal} {Science}\ }\textbf {\bibinfo {volume} {383}},\ \bibinfo {pages}
  {eadg8488} (\bibinfo {year} {2024})}\BibitemShut {NoStop}%
\bibitem [{\citenamefont
  {{Aguirre-L{\'o}pez}}(2024)}]{aguirre-lopez2024heterogeneous}%
  \BibitemOpen
  \bibfield  {author} {\bibinfo {author} {\bibfnamefont {F.}~\bibnamefont
  {{Aguirre-L{\'o}pez}}},\ }\href {https://doi.org/10.48550/arXiv.2404.11164}
  {\bibinfo {title} {Heterogeneous mean-field analysis of the generalized
  {{Lotka-Volterra}} model on a network}} (\bibinfo {year} {2024}),\ \Eprint
  {https://arxiv.org/abs/2404.11164} {arXiv:2404.11164 [cond-mat]} \BibitemShut
  {NoStop}%
\bibitem [{\citenamefont {Park}\ \emph {et~al.}(2024)\citenamefont {Park},
  \citenamefont {Lee}, \citenamefont {Lee},\ and\ \citenamefont
  {Park}}]{park2024incorporating}%
  \BibitemOpen
  \bibfield  {author} {\bibinfo {author} {\bibfnamefont {J.~I.}\ \bibnamefont
  {Park}}, \bibinfo {author} {\bibfnamefont {D.-S.}\ \bibnamefont {Lee}},
  \bibinfo {author} {\bibfnamefont {S.~H.}\ \bibnamefont {Lee}},\ and\ \bibinfo
  {author} {\bibfnamefont {H.~J.}\ \bibnamefont {Park}},\ }\href@noop {}
  {\bibinfo {title} {Incorporating {{Heterogeneous Interactions}} for
  {{Ecological Biodiversity}}}} (\bibinfo {year} {2024}),\ \Eprint
  {https://arxiv.org/abs/2403.15730} {arXiv:2403.15730 [cond-mat,
  physics:physics]} \BibitemShut {NoStop}%
\bibitem [{\citenamefont {Poley}\ \emph {et~al.}(2024)\citenamefont {Poley},
  \citenamefont {Galla},\ and\ \citenamefont {Baron}}]{poley2024interaction}%
  \BibitemOpen
  \bibfield  {author} {\bibinfo {author} {\bibfnamefont {L.}~\bibnamefont
  {Poley}}, \bibinfo {author} {\bibfnamefont {T.}~\bibnamefont {Galla}},\ and\
  \bibinfo {author} {\bibfnamefont {J.~W.}\ \bibnamefont {Baron}},\ }\href@noop
  {} {\bibinfo {title} {Interaction networks in persistent {{Lotka-Volterra}}
  communities}} (\bibinfo {year} {2024}),\ \Eprint
  {https://arxiv.org/abs/2404.08600} {arXiv:2404.08600 [cond-mat, q-bio]}
  \BibitemShut {NoStop}%
\bibitem [{\citenamefont {Turing}(1990)}]{turing1990chemical}%
  \BibitemOpen
  \bibfield  {author} {\bibinfo {author} {\bibfnamefont {A.~M.}\ \bibnamefont
  {Turing}},\ }\bibfield  {title} {\bibinfo {title} {The chemical basis of
  morphogenesis},\ }\href {https://doi.org/10.1007/BF02459572} {\bibfield
  {journal} {\bibinfo  {journal} {Bltn Mathcal Biology}\ }\textbf {\bibinfo
  {volume} {52}},\ \bibinfo {pages} {153} (\bibinfo {year} {1990})}\BibitemShut
  {NoStop}%
\bibitem [{\citenamefont {Dunne}\ \emph {et~al.}(2002)\citenamefont {Dunne},
  \citenamefont {Williams},\ and\ \citenamefont {Martinez}}]{dunne2002foodweb}%
  \BibitemOpen
  \bibfield  {author} {\bibinfo {author} {\bibfnamefont {J.~A.}\ \bibnamefont
  {Dunne}}, \bibinfo {author} {\bibfnamefont {R.~J.}\ \bibnamefont
  {Williams}},\ and\ \bibinfo {author} {\bibfnamefont {N.~D.}\ \bibnamefont
  {Martinez}},\ }\bibfield  {title} {\bibinfo {title} {Food-web structure and
  network theory: {{The}} role of connectance and size},\ }\href
  {https://doi.org/10.1073/pnas.192407699} {\bibfield  {journal} {\bibinfo
  {journal} {Proceedings of the National Academy of Sciences}\ }\textbf
  {\bibinfo {volume} {99}},\ \bibinfo {pages} {12917} (\bibinfo {year}
  {2002})}\BibitemShut {NoStop}%
\bibitem [{\citenamefont {Allesina}\ and\ \citenamefont
  {Pascual}(2009)}]{allesina2009food}%
  \BibitemOpen
  \bibfield  {author} {\bibinfo {author} {\bibfnamefont {S.}~\bibnamefont
  {Allesina}}\ and\ \bibinfo {author} {\bibfnamefont {M.}~\bibnamefont
  {Pascual}},\ }\bibfield  {title} {\bibinfo {title} {Food web models: A plea
  for groups},\ }\href {https://doi.org/10.1111/j.1461-0248.2009.01321.x}
  {\bibfield  {journal} {\bibinfo  {journal} {Ecology Letters}\ }\textbf
  {\bibinfo {volume} {12}},\ \bibinfo {pages} {652} (\bibinfo {year}
  {2009})}\BibitemShut {NoStop}%
\bibitem [{\citenamefont {Allesina}\ \emph {et~al.}(2015)\citenamefont
  {Allesina}, \citenamefont {Grilli}, \citenamefont {Barab{\'a}s},
  \citenamefont {Tang}, \citenamefont {Aljadeff},\ and\ \citenamefont
  {Maritan}}]{allesina2015predicting}%
  \BibitemOpen
  \bibfield  {author} {\bibinfo {author} {\bibfnamefont {S.}~\bibnamefont
  {Allesina}}, \bibinfo {author} {\bibfnamefont {J.}~\bibnamefont {Grilli}},
  \bibinfo {author} {\bibfnamefont {G.}~\bibnamefont {Barab{\'a}s}}, \bibinfo
  {author} {\bibfnamefont {S.}~\bibnamefont {Tang}}, \bibinfo {author}
  {\bibfnamefont {J.}~\bibnamefont {Aljadeff}},\ and\ \bibinfo {author}
  {\bibfnamefont {A.}~\bibnamefont {Maritan}},\ }\bibfield  {title} {\bibinfo
  {title} {Predicting the stability of large structured food webs},\ }\href
  {https://doi.org/10.1038/ncomms8842} {\bibfield  {journal} {\bibinfo
  {journal} {Nat Commun}\ }\textbf {\bibinfo {volume} {6}},\ \bibinfo {pages}
  {7842} (\bibinfo {year} {2015})}\BibitemShut {NoStop}%
\bibitem [{\citenamefont {Poley}\ \emph
  {et~al.}(2023{\natexlab{a}})\citenamefont {Poley}, \citenamefont {Baron},\
  and\ \citenamefont {Galla}}]{poley2023generalized}%
  \BibitemOpen
  \bibfield  {author} {\bibinfo {author} {\bibfnamefont {L.}~\bibnamefont
  {Poley}}, \bibinfo {author} {\bibfnamefont {J.~W.}\ \bibnamefont {Baron}},\
  and\ \bibinfo {author} {\bibfnamefont {T.}~\bibnamefont {Galla}},\ }\bibfield
   {title} {\bibinfo {title} {Generalized {{Lotka-Volterra}} model with
  hierarchical interactions},\ }\href
  {https://doi.org/10.1103/PhysRevE.107.024313} {\bibfield  {journal} {\bibinfo
   {journal} {Phys. Rev. E}\ }\textbf {\bibinfo {volume} {107}},\ \bibinfo
  {pages} {024313} (\bibinfo {year} {2023}{\natexlab{a}})}\BibitemShut
  {NoStop}%
\bibitem [{\citenamefont {Poley}\ \emph
  {et~al.}(2023{\natexlab{b}})\citenamefont {Poley}, \citenamefont {Galla},\
  and\ \citenamefont {Baron}}]{poley2023eigenvalue}%
  \BibitemOpen
  \bibfield  {author} {\bibinfo {author} {\bibfnamefont {L.}~\bibnamefont
  {Poley}}, \bibinfo {author} {\bibfnamefont {T.}~\bibnamefont {Galla}},\ and\
  \bibinfo {author} {\bibfnamefont {J.~W.}\ \bibnamefont {Baron}},\ }\href@noop
  {} {\bibinfo {title} {Eigenvalue spectra of finely structured random
  matrices}} (\bibinfo {year} {2023}{\natexlab{b}}),\ \Eprint
  {https://arxiv.org/abs/2311.02006} {arXiv:2311.02006 [cond-mat, q-bio]}
  \BibitemShut {NoStop}%
\bibitem [{\citenamefont {Crooks}\ \emph {et~al.}(2011)\citenamefont {Crooks},
  \citenamefont {Burdett}, \citenamefont {Theobald}, \citenamefont
  {Rondinini},\ and\ \citenamefont {Boitani}}]{crooks2011global}%
  \BibitemOpen
  \bibfield  {author} {\bibinfo {author} {\bibfnamefont {K.~R.}\ \bibnamefont
  {Crooks}}, \bibinfo {author} {\bibfnamefont {C.~L.}\ \bibnamefont {Burdett}},
  \bibinfo {author} {\bibfnamefont {D.~M.}\ \bibnamefont {Theobald}}, \bibinfo
  {author} {\bibfnamefont {C.}~\bibnamefont {Rondinini}},\ and\ \bibinfo
  {author} {\bibfnamefont {L.}~\bibnamefont {Boitani}},\ }\bibfield  {title}
  {\bibinfo {title} {Global patterns of fragmentation and connectivity of
  mammalian carnivore habitat},\ }\href
  {https://doi.org/10.1098/rstb.2011.0120} {\bibfield  {journal} {\bibinfo
  {journal} {Philos Trans R Soc Lond B Biol Sci}\ }\textbf {\bibinfo {volume}
  {366}},\ \bibinfo {pages} {2642} (\bibinfo {year} {2011})}\BibitemShut
  {NoStop}%
\bibitem [{\citenamefont {Crooks}\ \emph {et~al.}(2017)\citenamefont {Crooks},
  \citenamefont {Burdett}, \citenamefont {Theobald}, \citenamefont {King},
  \citenamefont {Di~Marco}, \citenamefont {Rondinini},\ and\ \citenamefont
  {Boitani}}]{crooks2017quantification}%
  \BibitemOpen
  \bibfield  {author} {\bibinfo {author} {\bibfnamefont {K.~R.}\ \bibnamefont
  {Crooks}}, \bibinfo {author} {\bibfnamefont {C.~L.}\ \bibnamefont {Burdett}},
  \bibinfo {author} {\bibfnamefont {D.~M.}\ \bibnamefont {Theobald}}, \bibinfo
  {author} {\bibfnamefont {S.~R.~B.}\ \bibnamefont {King}}, \bibinfo {author}
  {\bibfnamefont {M.}~\bibnamefont {Di~Marco}}, \bibinfo {author}
  {\bibfnamefont {C.}~\bibnamefont {Rondinini}},\ and\ \bibinfo {author}
  {\bibfnamefont {L.}~\bibnamefont {Boitani}},\ }\bibfield  {title} {\bibinfo
  {title} {Quantification of habitat fragmentation reveals extinction risk in
  terrestrial mammals},\ }\href {https://doi.org/10.1073/pnas.1705769114}
  {\bibfield  {journal} {\bibinfo  {journal} {Proc. Natl. Acad. Sci. U.S.A.}\
  }\textbf {\bibinfo {volume} {114}},\ \bibinfo {pages} {7635} (\bibinfo {year}
  {2017})}\BibitemShut {NoStop}%
\bibitem [{\citenamefont {Yonatan}\ \emph {et~al.}(2022)\citenamefont
  {Yonatan}, \citenamefont {Amit}, \citenamefont {Friedman},\ and\
  \citenamefont {Bashan}}]{yonatan2022complexity}%
  \BibitemOpen
  \bibfield  {author} {\bibinfo {author} {\bibfnamefont {Y.}~\bibnamefont
  {Yonatan}}, \bibinfo {author} {\bibfnamefont {G.}~\bibnamefont {Amit}},
  \bibinfo {author} {\bibfnamefont {J.}~\bibnamefont {Friedman}},\ and\
  \bibinfo {author} {\bibfnamefont {A.}~\bibnamefont {Bashan}},\ }\bibfield
  {title} {\bibinfo {title} {Complexity--stability trade-off in empirical
  microbial ecosystems},\ }\href {https://doi.org/10.1038/s41559-022-01745-8}
  {\bibfield  {journal} {\bibinfo  {journal} {Nat Ecol Evol}\ }\textbf
  {\bibinfo {volume} {6}},\ \bibinfo {pages} {693} (\bibinfo {year}
  {2022})}\BibitemShut {NoStop}%
\bibitem [{\citenamefont {Hu}\ \emph {et~al.}(2022)\citenamefont {Hu},
  \citenamefont {Amor}, \citenamefont {Barbier}, \citenamefont {Bunin},\ and\
  \citenamefont {Gore}}]{hu2022emergent}%
  \BibitemOpen
  \bibfield  {author} {\bibinfo {author} {\bibfnamefont {J.}~\bibnamefont
  {Hu}}, \bibinfo {author} {\bibfnamefont {D.~R.}\ \bibnamefont {Amor}},
  \bibinfo {author} {\bibfnamefont {M.}~\bibnamefont {Barbier}}, \bibinfo
  {author} {\bibfnamefont {G.}~\bibnamefont {Bunin}},\ and\ \bibinfo {author}
  {\bibfnamefont {J.}~\bibnamefont {Gore}},\ }\bibfield  {title} {\bibinfo
  {title} {Emergent phases of ecological diversity and dynamics mapped in
  microcosms},\ }\href {https://doi.org/10.1126/science.abm7841} {\bibfield
  {journal} {\bibinfo  {journal} {Science}\ }\textbf {\bibinfo {volume}
  {378}},\ \bibinfo {pages} {85} (\bibinfo {year} {2022})}\BibitemShut
  {NoStop}%
\bibitem [{\citenamefont {Tang}\ and\ \citenamefont
  {Allesina}(2014)}]{tang2014reactivity}%
  \BibitemOpen
  \bibfield  {author} {\bibinfo {author} {\bibfnamefont {S.}~\bibnamefont
  {Tang}}\ and\ \bibinfo {author} {\bibfnamefont {S.}~\bibnamefont
  {Allesina}},\ }\bibfield  {title} {\bibinfo {title} {Reactivity and stability
  of large ecosystems},\ }\href@noop {} {\bibfield  {journal} {\bibinfo
  {journal} {Frontiers in Ecology and Evolution}\ }\textbf {\bibinfo {volume}
  {2}} (\bibinfo {year} {2014})}\BibitemShut {NoStop}%
\bibitem [{\citenamefont {Arnoldi}(1951)}]{arnoldi1951principle}%
  \BibitemOpen
  \bibfield  {author} {\bibinfo {author} {\bibfnamefont {W.~E.}\ \bibnamefont
  {Arnoldi}},\ }\bibfield  {title} {\bibinfo {title} {The principle of
  minimized iterations in the solution of the matrix eigenvalue problem},\
  }\href {https://doi.org/10.1090/qam/42792} {\bibfield  {journal} {\bibinfo
  {journal} {Quart. Appl. Math.}\ }\textbf {\bibinfo {volume} {9}},\ \bibinfo
  {pages} {17} (\bibinfo {year} {1951})}\BibitemShut {NoStop}%
\bibitem [{\citenamefont {Janson}(2018)}]{janson2018edge}%
  \BibitemOpen
  \bibfield  {author} {\bibinfo {author} {\bibfnamefont {S.}~\bibnamefont
  {Janson}},\ }\bibfield  {title} {\bibinfo {title} {On {{Edge Exchangeable
  Random Graphs}}},\ }\href {https://doi.org/10.1007/s10955-017-1832-9}
  {\bibfield  {journal} {\bibinfo  {journal} {J Stat Phys}\ }\textbf {\bibinfo
  {volume} {173}},\ \bibinfo {pages} {448} (\bibinfo {year}
  {2018})}\BibitemShut {NoStop}%
\bibitem [{\citenamefont {Newman}\ \emph {et~al.}(2001)\citenamefont {Newman},
  \citenamefont {Strogatz},\ and\ \citenamefont {Watts}}]{newman2001random}%
  \BibitemOpen
  \bibfield  {author} {\bibinfo {author} {\bibfnamefont {M.~E.~J.}\
  \bibnamefont {Newman}}, \bibinfo {author} {\bibfnamefont {S.~H.}\
  \bibnamefont {Strogatz}},\ and\ \bibinfo {author} {\bibfnamefont {D.~J.}\
  \bibnamefont {Watts}},\ }\bibfield  {title} {\bibinfo {title} {Random graphs
  with arbitrary degree distributions and their applications},\ }\href
  {https://doi.org/10.1103/PhysRevE.64.026118} {\bibfield  {journal} {\bibinfo
  {journal} {Phys. Rev. E}\ }\textbf {\bibinfo {volume} {64}},\ \bibinfo
  {pages} {026118} (\bibinfo {year} {2001})}\BibitemShut {NoStop}%
\bibitem [{\citenamefont {Newman}(2003)}]{newman2003properties}%
  \BibitemOpen
  \bibfield  {author} {\bibinfo {author} {\bibfnamefont {M.~E.~J.}\
  \bibnamefont {Newman}},\ }\bibfield  {title} {\bibinfo {title} {Properties of
  highly clustered networks},\ }\href
  {https://doi.org/10.1103/PhysRevE.68.026121} {\bibfield  {journal} {\bibinfo
  {journal} {Phys. Rev. E}\ }\textbf {\bibinfo {volume} {68}},\ \bibinfo
  {pages} {026121} (\bibinfo {year} {2003})}\BibitemShut {NoStop}%
\bibitem [{\citenamefont {Newman}\ and\ \citenamefont
  {Watts}(1999)}]{newman1999scaling}%
  \BibitemOpen
  \bibfield  {author} {\bibinfo {author} {\bibfnamefont {M.~E.~J.}\
  \bibnamefont {Newman}}\ and\ \bibinfo {author} {\bibfnamefont {D.~J.}\
  \bibnamefont {Watts}},\ }\bibfield  {title} {\bibinfo {title} {Scaling and
  percolation in the small-world network model},\ }\href
  {https://doi.org/10.1103/PhysRevE.60.7332} {\bibfield  {journal} {\bibinfo
  {journal} {Phys. Rev. E}\ }\textbf {\bibinfo {volume} {60}},\ \bibinfo
  {pages} {7332} (\bibinfo {year} {1999})}\BibitemShut {NoStop}%
\bibitem [{\citenamefont {Fall}\ \emph {et~al.}(2007)\citenamefont {Fall},
  \citenamefont {Fortin}, \citenamefont {Manseau},\ and\ \citenamefont
  {O'Brien}}]{fall2007spatial}%
  \BibitemOpen
  \bibfield  {author} {\bibinfo {author} {\bibfnamefont {A.}~\bibnamefont
  {Fall}}, \bibinfo {author} {\bibfnamefont {M.-J.}\ \bibnamefont {Fortin}},
  \bibinfo {author} {\bibfnamefont {M.}~\bibnamefont {Manseau}},\ and\ \bibinfo
  {author} {\bibfnamefont {D.}~\bibnamefont {O'Brien}},\ }\bibfield  {title}
  {\bibinfo {title} {Spatial {{Graphs}}: {{Principles}} and {{Applications}}
  for {{Habitat Connectivity}}},\ }\href
  {https://doi.org/10.1007/s10021-007-9038-7} {\bibfield  {journal} {\bibinfo
  {journal} {Ecosystems}\ }\textbf {\bibinfo {volume} {10}},\ \bibinfo {pages}
  {448} (\bibinfo {year} {2007})}\BibitemShut {NoStop}%
\bibitem [{\citenamefont {Falkenberg}\ \emph {et~al.}(2020)\citenamefont
  {Falkenberg}, \citenamefont {Lee}, \citenamefont {Amano}, \citenamefont
  {Ogawa}, \citenamefont {Yano}, \citenamefont {Miyake}, \citenamefont
  {Evans},\ and\ \citenamefont {Christensen}}]{falkenberg2020identifying}%
  \BibitemOpen
  \bibfield  {author} {\bibinfo {author} {\bibfnamefont {M.}~\bibnamefont
  {Falkenberg}}, \bibinfo {author} {\bibfnamefont {J.-H.}\ \bibnamefont {Lee}},
  \bibinfo {author} {\bibfnamefont {S.-i.}\ \bibnamefont {Amano}}, \bibinfo
  {author} {\bibfnamefont {K.-i.}\ \bibnamefont {Ogawa}}, \bibinfo {author}
  {\bibfnamefont {K.}~\bibnamefont {Yano}}, \bibinfo {author} {\bibfnamefont
  {Y.}~\bibnamefont {Miyake}}, \bibinfo {author} {\bibfnamefont {T.~S.}\
  \bibnamefont {Evans}},\ and\ \bibinfo {author} {\bibfnamefont
  {K.}~\bibnamefont {Christensen}},\ }\bibfield  {title} {\bibinfo {title}
  {Identifying time dependence in network growth},\ }\href
  {https://doi.org/10.1103/PhysRevResearch.2.023352} {\bibfield  {journal}
  {\bibinfo  {journal} {Phys. Rev. Research}\ }\textbf {\bibinfo {volume}
  {2}},\ \bibinfo {pages} {023352} (\bibinfo {year} {2020})}\BibitemShut
  {NoStop}%
\bibitem [{\citenamefont {Barter}\ and\ \citenamefont
  {Gross}(2017)}]{barter2017spatial}%
  \BibitemOpen
  \bibfield  {author} {\bibinfo {author} {\bibfnamefont {E.}~\bibnamefont
  {Barter}}\ and\ \bibinfo {author} {\bibfnamefont {T.}~\bibnamefont {Gross}},\
  }\bibfield  {title} {\bibinfo {title} {Spatial effects in meta-foodwebs},\
  }\href {https://doi.org/10.1038/s41598-017-08666-8} {\bibfield  {journal}
  {\bibinfo  {journal} {Sci Rep}\ }\textbf {\bibinfo {volume} {7}},\ \bibinfo
  {pages} {9980} (\bibinfo {year} {2017})}\BibitemShut {NoStop}%
\end{thebibliography}%

\clearpage
\section*{Supplementary materials}\label{sec:supp}%
\supplementary%

\subsection{Generalized Lotka-Volterra model for meta-ecosystems}\label[supp]{sec:SI:glv}%
We consider a meta-ecosystem with $S$ species and $M$ patches.
While we study a linearized model, the elements of the community matrix of~\cref{eq:J} originate from the description of a dynamical system.
Local dynamics of all species $\vb{N}^\mu = (N_1^\mu, \ldots, N_S^\mu)$, that $N_i^\mu$ denotes the abundance of species $i$ on patch $\mu$, are governed by a generalized Lotka-Volterra model that includes dispersal between adjacent patches $\nu$, and is given by
\begin{align}
  \dv{N_i^\mu}{t} &= N_i^\mu \left( r_i^\mu - \kappa_i^\mu N_i^\mu - \sum_{j\neq i} A_{ij}^\mu N_j^\mu \right)
  + \Delta N_i^\mu, \\
  &\qq{where} \Delta N_i^\mu = \sum_\nu f_i^{\mu\nu}(\vb{N}^\mu, \vb{N}^\nu), \nonumber
\end{align}
where $r_i^\mu$ the growth rate of species $i$ on patch $\mu$, $A_{ij}^\mu$ the interaction coefficient, $\kappa_i^\mu$ the self-interaction (related to the carrying capacity), and $f_i^{\mu\nu}$ a general (non-linear) density dependent dispersal function between (adjacent) patches $\mu$ and $\nu$.

\subsubsection{The Jacobian and the community matrix}\label[supp]{sec:SI:communitymatrix}%
To proceed, we specify the dispersal function as
\begin{align}
  f_i^{\mu\nu}(\vb{N}^\mu,\vb{N}^\nu) = \gamma_i^{\mu\nu} (N_i^\nu - N_i^\mu),
\end{align}
and derive the non-zero elements of the Jacobian matrix $\vb*{\mathfrak{J}}$, which read
\begin{align}
  \begin{split}
    \mathfrak{J}_{ii}^{\mu\nu} &= \gamma_i^{\mu\nu} \\
    \mathfrak{J}_{ij}^{\mu\mu} &= -A_{ij} N_i^\mu \\
    \mathfrak{J}_{ii}^{\mu\mu} &= r_i^\mu - \sum_j A_{ij}^\mu N_j ^\mu
                                 - \sum_\nu \gamma_i^{\mu\nu},
  \end{split}
\end{align}
where in the last term we have absorbed the $\kappa_i^\mu N_i^\mu$-terms in the $A_{ii}^\mu$ elements.
The Jacobian is block-structured (see~\cref{fig:metaecosystem,fig:blockcommunity}) with diagonal blocks consisting of; (i)~diagonal matrices with (species-specific) growth rates on its diagonal, (ii)~diagonal matrices with the total (outgoing) dispersal rate to adjacent patches on its diagonal, and (iii)~local interaction matrices.
Its off-diagonal blocks are themselves diagonal matrices with (incoming) dispersal rates on its diagonal.
As such, the community matrix $\vb{J}$ --- which is the Jacobian $\vb*{\mathfrak{J}}$ evaluated at the (hypothetical) feasible fixed point with $(N_i^\mu)^\star > 0$ --- is also block-structured. 
Note that in the absence of dispersal one recovers the standard community matrix for the Lotka-Volterra model as for $\gamma_i^{\mu\nu} = 0$ we have, in the fixed point, $r_i^\mu - \sum_j A_{ij}^\mu N_j^\mu = 0$.

In the spirit of \citet{may1972will}, 
we restrict ourselves to the stationary and linear regime of the generalized Lotka-Volterra model.
We do this as to allow a direct comparison between our approach, which includes explicit spatial structure, and works that do not~(in particular, Refs.~\citep{gravel2016stability,baron2020dispersalinduced}, but see also~\citep{allesina2012stability,tang2014reactivity,allesina2015stability,allesina2020models}, among others).
This means that we consider the community matrix to be written as the sum of three matrices \citep{gravel2016stability,baron2020dispersalinduced} %
\begin{align}
  \vb{J} = \vb{R} + \vb{D} + \vb{A},
\end{align}
where $\vb{R}$, $\vb{D}$, and $\vb{D}$ correspond to descriptions (i), (ii), and (iii), evaluated at the feasible fixed point.

\subsubsection{The community matrix as a random block matrix}\label[supp]{sec:SI:blockmatrix}%
In order to reason about the influence of dispersal on the (linear) stability of the feasible fixed point, we formally introduce the network $\mathscr{G} = (\mathcal{V}, \mathcal{E})$ with the set of vertices $\mathcal{V}$ corresponding to the patches and the set of edges $\mathcal{E}$ that specify whether dispersal between them is possible.
We let $\vbm{G}$ be the adjacency matrix of the network $\mathscr{G}$ and write the per-capita dispersal rate between two patches $\mu$ and $\nu$ as
\begin{align}
  \gamma_i^{\mu\nu} = \gamma_i \mathcal{G}_{\mu\nu}.  
\end{align}
We assume dispersal does not lead to changes in abundances, so we let the \emph{dispersal matrix}, which is a block matrix, $\vb{D}$ be defined by
\begin{align}
  \label{eq:SI:D}
  D_{ii}^{\mu\nu} =
  \begin{cases}
    -\gamma_i &\qq{when} \mu = \nu, \\
    \gamma_i/k_\mu &\qq{when} \mathcal{G}_{\mu\nu} = 1,
  \end{cases}
\end{align}
where $k_\mu$ is the \emph{degree} of vertex $\mu$, i.e.~the number of patches connected to patch $\mu$, i.e. $k_\mu = \sum_\nu \mathcal{G}_{\mu\nu}$.
As we shall consider networks $\mathscr{G}$ to be random networks (i.e., the network is generated using \emph{some} random process), $\vb{D}$ is generally a random (block) matrix.

Next we write the blocked local interaction matrix as $\vb{A} = -a\vb{I} + \vb{B}$, with $a$ the self-regulation strength, $\vb{I}$ the identity matrix, and $\vb{B}$ a random block-diagonal matrix.
Diagonal blocks, denoted with ${\vb{B}}_{\mu} \equiv {(\vb{B})}_{\mu\mu}$, have random elements with mean $\langle b_{ij}^\mu \rangle = 0$, variance $\langle {(b_{ij}^\mu)}^2 \rangle = c\sigma^2$, and between-patch correlations $\langle b_{ij}^\mu b_{ij}^\nu \rangle = \rho c\sigma^2/S$.
Here, using standard conventions (see, e.g.~\citep{may1972will,allesina2012stability,allesina2015stability}, among others), we have introduced the \emph{connectance} $c$ that defines the probability of a pairwise interaction occurring, i.e.~with probability $c$ elements $b_{ij}^\mu = 0$, and with probability $1-c$ they are sampled from a distribution with the above-mentioned statistics.
Finally, diagonal elements $b_{ii}^\mu$ are set to $0$.

As both the dispersal matrix and the interaction matrix are random matrices (albeit from vastly different random processes), the community matrix is a random matrix as well.
We are interested in the linear stability of the feasible fixed point, thus we are interested in obtaining either the full eigenvalue spectral distribution (ESD), $p(\lambda)$, or, at least, the (average) largest right-most eigenvalue
\begin{align}
  \label{eq:lambda1}
  \lambda_1 = \max_i \Re \lambda_i(\vb{J}),
\end{align}
where $\lambda_i(\vb{J})$ are the eigenvalues of $\vb{J}$.
In some cases (see below), one can obtain a closed-form solution of the distribution of eigenvalues.
However, in general, determination of the distributions is difficult as $\vb{J}$ contains sums of matrices that are generated with vastly different random processes.

While perturbation-based methods might appear fruitful to obtain approximations of the eigenvalues, the regime wherein these methods hold is biologically uninteresting.
More formally, perturbative methods essentially work by stating that the eigenvalues will be shifted by some small amount, i.e.~$\lambda^\prime_i \approx \lambda_i + \delta \lambda_i$, where $\delta \lambda_i$ is the perturbation and thus originates from one of the matrices, here either the interaction matrix $\vb{B}$ or the dispersal matrix $\vb{D}$.
However, for this to approximate the distribution of $\vb{J}$ reasonably well, one needs either that $||\vb{D}|| \ll ||\vb{B}||$ (low dispersal regime) or vice versa (high dispersal regime), with $||\cdot||$ some norm, such as the matrix norm.
Clearly, the low dispersal regime essentially assumes dispersal to be absent, in which one obviously cannot study the effects of dispersal.
In the high dispersal regime, interactions between species need to be extremely weak, essentially omitting the effect of species interactions entirely.
As such, the regimes wherein perturbation-bases methods hold are biologically uninteresting and therefore we instead resort here to numerical determination of the (right-most) eigenvalues.

\subsection{Numerical computation of eigenvalues}\label[supp]{sec:SI:numericalesd}%
Depending on the results presented, we compute the eigenvalues of randomly generated community matrices using \texttt{Julia}.
When the number of species $S$ and the number of patches $M$ are both large, computation of all eigenvalues is computationally costly.
Hence, we resort to computing only a few eigenvalues using Arnoldi iteration~\citep{arnoldi1951principle}, which is implemented in \texttt{ArnoldiMethod.jl}\footnote{\url{https://github.com/JuliaLinearAlgebra/ArnoldiMethod.jl}}, which allows us to study stability of systems that comprise many species and a (relatively) large number of patches.
Code to produce the presented results is available upon request.

\subsection{The eigenvalue spectrum of meta-ecosystems}\label[supp]{sec:SI:esd}%
In order to obtain (an approximate) description of $\lambda_1$, we shall first introduce recently obtained results in fully-connected systems~\citep{gravel2016stability,baron2020dispersalinduced}.
As the results from Ref.~\citep{baron2020dispersalinduced} generalize those of Ref.~\citep{gravel2016stability}, we shall here summarize the results relevant for the systems considered here.

We note that the descriptions in Ref.~\citep{baron2020dispersalinduced} are more general with respect to interaction structure, but only local dispersal (on cycle-, or ring-networks) and all-to-all dispersal (on fully-connected) networks were considered. 
Nevertheless, when the networks are fully-connected, an inequality for the support of the eigenvalue spectrum can be obtained (see Ref.~\citep[p. S53]{baron2020dispersalinduced}), which for $\rho = 0$ reads
\begin{align}
  \label{eq:supportesdbaron}
  \frac{1}{|z+m|^2} + \frac{M - 1}{|z + m + \gamma M / (M-1) |^2} &\leq \frac{M}{cS\sigma^2},
\end{align}
where $m = b - r$.
The stability criterion can be obtained from the support by looking at its boundary for $z$ real (i.e., $\Im z = 0$), and solving for equality.
For $\gamma$ large, one can obtain an approximation of the criterion at first order in $1/\gamma$, which reads,
\begin{align}
  \label{eq:SI:stabilitycriterionlargegamma}
  \sigma\sqrt{cS/M} < m.
\end{align}
For small $\gamma$, we can instead obtain a first order approximation in $\gamma$, for which the stability criterion becomes,
\begin{align}
  \sigma\sqrt{cS} < m + \gamma.
\end{align}
These are the criteria mentioned in the main body (\cref{eq:stabilitycriterionsmallgamma,eq:stabilitycriterionlarge}).
Note that a (more lengthy) expression that holds for all values of $\rho$ is also available, but we omit it here for brevity and refer the interested reader to Ref.~\citep{baron2020dispersalinduced}.
For low edge densities, however, \cref{eq:supportesdbaron} does not hold as the specific network topology significantly alters the support of the eigenvalue spectrum (see~\cref{fig:LRevpoisson,fig:LRevwattsstrogatz,fig:rggphaseplot} and~\cref{fig:SI:LRevrgg}).

\begin{figure}[t]
  \centering
  \includegraphics[width=.7\columnwidth]{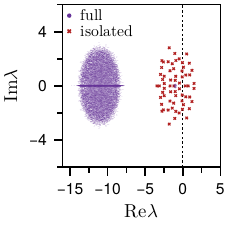}%
  \caption{%
    Addition of a single isolated patch can destabilize an entire system when $\gamma \gg 1$.
    All eigenvalues of a fully connected system are shown in purple, while red crosses are the eigenvalues of a single isolated patch (vertex).
    Note that some of the eigenvalues of the isolated patch have positive real value, thus implying linear instability.
    Additionally note that $S$ eigenvalues near centered at $r-b=-1$ with (small) radius $\sigma\sqrt{cS/M}$.
    Relevant parameters are $S=64$, $M_0=256$, $c=0.3$, $\sigma=1/\sqrt{S}$, $\rho=0$, and $\gamma=10$.
  }%
  \label{fig:SI:isolated}
\end{figure}

\subsection{Isolated patches and stability}\label[supp]{sec:SI:isolated}%
To illustrate the effect of isolated patches on stability, consider a simple example of a connected system with $M_0$ patches and $S$ species.
Then, introduce a single isolated patch such that $M = M_0+1$, and note that the eigenvalues of the community matrix $\vb{J}$ are, in this case, the union of the spectral distributions of the connected system and of the isolated subsystem.
The reason is that none of the block-rows nor block-columns at the $M$th index have non-zero value --- only the diagonal block is a non-zero matrix, and is equal to the interaction matrix on the newly introduced isolated patch.
Following standard arguments we know that the eigenvalues of the isolated patch are within a circle centered at $(r-b,0)$ with radius $\sigma\sqrt{cS}$, which is the classical stability criterion of \citet{may1972will}.
Thus, when the isolated patch is locally unstable, the entire system is unstable (\cref{fig:SI:isolated}).
When we connect the isolated patch to the large connected component, the largest eigenvalue of the spectrum is still centered at $(r-b,0)$, but now the radius shrinks proportional to $\sqrt{1/M}$ (for $\gamma$ large, see~\cref{eq:SI:stabilitycriterionlargegamma}, \cref{fig:SI:isolated}), and thus the system is more likely to be stable.

\subsection{The configuration model and giant components}\label[supp]{sec:SI:configmodel}%
As the above example illustrated, we are interested not in the effect of network connectivity on stability, but of network interconnectivity.
Hence, we focus on the degree distribution of giant (connected) components.
To facilitate this, we first consider dispersal networks to be generated using the configuration model \citep{newman2018networks}.
A configuration model network is a network whose degree sequence is sampled from an arbitrary degree distribution, $p(k)$.
Typically configuration model networks do not exhibit degree-degree correlations, meaning that the local structure is tree-like.

To generate a network, one samples a degree sequence $\vb{k} = (k_1, \ldots, k_M)$ independently from $p(k)$.
In practice, node degrees $k_i$ are often subjected to bounds such that $k_{\min} \leq k_i \leq k_{\max}$, and specific choices of $k_{\min}$ and $k_{\max}$ changes network characteristics.
In our particular case, $k_{\min} \geq 1$ as we wish to avoid isolated vertices, and of course $k_{\max} \leq M-1$.
When $k_{\min} = 1$, the configuration networks exhibit three distinct phases \citep{newman2018networks,tishby2019generating}: (i)~a sparse limit, where no giant component exists, (ii)~a dense limit, where all vertices belong to a single giant component, and (iii)~and intermediate regime, above the percolation threshold, where a giant component and finite tree elements coexist.
As we are interested in connected, but not necessarily dense networks, we are interested in both the intermediate regime and the dense regime.

\subsubsection{The degree distribution and the giant component}\label[supp]{sec:SI:degreedistribution}%
Recently, progress has been made on descriptions of the degree distribution of the giant component, which has been shown to differ significantly from the global degree distribution \citep{tishby2018revealing,tishby2018statistical}.
To obtain the degree distribution of the giant component, we define the degree distribution of the full network with $p_0(k)$.
The degree distribution of the giant component and of the finite components are denoted with $p(k)$ and $q(k)$ respectively. 
Then, denote with $g$ the probability that a random node $i$ belongs to the giant component, and $h$ the probability that a random neighbor of $i$ belongs to the giant component of the reduced network that does not include $i$.
Then, the degree distribution, conditioned on the giant component, has been shown to be given by \citep{tishby2018revealing,tishby2019generating}
\begin{align}
  \label{eq:pkconfig}
  p(k) = \frac{1 - {(1-h)}^k}{g} p_0(k),
\end{align}
which has mean degree
\begin{align}
  \label{eq:ekconfig}
  \mathbb{E}[k] = \frac{1-{(1-h)}^2}{g} \mathbb{E}_0[k],
\end{align}
where $\mathbb{E}_0[k] = \sum_k k p_0(k)$ the mean degree of the full network.
Next, we wish that the expected size of the giant component equals $M$, which enables us to control for network size.
In general, the expected value of the size of the giant component of a configuration network is given by $\langle M \rangle = g M_0$, where $M_0$ the number of nodes of the full network.

The goal then becomes to choose a suitable degree distribution $p_0(k)$ such that the degree distribution of the giant component $p(k)$, and the number of nodes $M$, is as desired.
Inverting \cref{eq:pkconfig} one obtains 
\begin{align}
  \label{eq:p0kconfig}
  p_0(k) = \frac{g}{1 - {(1 - h)}^k} p(k),
\end{align}
and similarly for the number of nodes $M_0$ we obtain
\begin{align}
  M_0 = \langle M \rangle / g.
\end{align}
Thus, to obtain a connected network of size $\langle M \rangle$ with degree distribution $p(k)$, one needs to generate networks of size $M_0$ and degree distribution $p_0(k)$.
The giant component of these networks is then the desired single-component network.

To do this, introduce the generating functions
\begin{align}
  G(x) = \sum_{k=1}^\infty x^k p(k), \qquad
  H(x) = \sum_{k=1}^\infty \frac{kx^{(k-1)}}{\mathbb{E}[k]} p(k).
\end{align}
In order to compute $h$, we use \cref{eq:pkconfig,eq:ekconfig}, and obtain \citep{tishby2019generating}
\begin{align}
  {(1-h)}^2 &= 1 - \frac{1}{
              \sum_{n=0}^\infty {(1-h)}^n H[{(1-h)}^n]
              }.
\end{align}
This implicit equation can be approximated under some conditions, which leads to the equation
\begin{align}
  h(2-h) \left[ 1 + (1-h)H(1-h) - \frac{G[{(1-h)}^{3/2}]}{\mathbb{E}[k] \log (1-h)} \right] = 1,
\end{align}
which usually has to be solved numerically.
Once $h$ has been calculated it can be used to obtain $g$,
\begin{align}
  g = {\left( 1 + \sum_{n=1}^\infty G[{(1-h)}^n] \right)}^{-1},
\end{align}
which can, in turn, be used to compute $p_0(k)$ using \cref{eq:p0kconfig}.

Note that the limits of the sum consider that $k$ can be large (or rather, $M\rightarrow\infty$).
In practice, the degree distribution is limited by the size of the network, e.g. $k \leq k_{\max} = M-1$, which, for $M$ large but finite, can be much lower depending on the distribution.
As such, the sums can be approximated numerically, simply by considering terms up to $k_{\max}$ or stopping when contributions to the generating function become negligible\footnote{For example, one can limit contributions to the systems' precision, but we empirically established that some orders of magnitude above that suffice.}.

Finally note that is it also possible to control the exact size of the giant component \citep{tishby2019generating}, instead of having the giant component be a random variable $\widetilde{M}$ with mean $\langle M \rangle$.
For details on the procedure of adding or removing vertices, depending on whether $\widetilde{M}$ is larger of smaller than (the integer part of) $\langle M \rangle$, we refer the interested reader to \citet[p.5]{tishby2019generating}.

To summarize, the scheme above allows us to generate \emph{connected} networks with any arbitrary degree distribution.
As such, we avoid the problem of introducing patches which are disconnected from the system that could potentially destabilize the system.
We will now illustrate the procedure with some examples of networks with a common degree distribution.

\subsubsection{Networks where the giant component has a Poisson degree distribution}\label[supp]{sec:SI:poisson}%

\begin{figure}[t]
  \centering
  \includegraphics[width=.7\columnwidth]{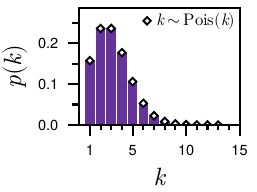}%
  \caption{%
    Degree distribution of connected components of $10^4$ networks sampled using $p_0(k)$ given by \cref{eq:p0kconfig} and $p(k)$ a Poisson distribution with $M=512$, $\langle k \rangle = 3$ and $k_{\min}=1$.
    Bars represent numerically obtained histogram and markers indicate the desired distribution given by~\cref{eq:SI:poisson}.
  }%
  \label{fig:SI:pkER}
\end{figure}

To highlight the difference between the degree distribution of the full network --- i.e.~the network that included both the giant component (if it exists) and all the finite components --- let us initially consider \emph{Poisson networks}, i.e.~networks of which the degree distribution of the giant component is a Poisson distribution for all degrees $k\geq k_{\min} = 1$,
\begin{align}
  \label{eq:SI:poisson}
  p(k) = \frac{e^{-\langle k \rangle}\langle k \rangle^k}{(1-e^{-\langle k \rangle}) k!}
\end{align}
meaning that the resulting network could be considered an Erd\H{o}s-R\'enyi network with the additional constraints that $k \geq k_{\min} = 1$ and mean degree $\langle k \rangle \geq 2$.
To illustrate the results of the procedure of~\citet{tishby2019generating}, we have plotted numerically obtained degree distributions and compared these with~\cref{eq:SI:poisson} in~\cref{fig:SI:pkER}.
One can appreciate that we can indeed sample giant components with the appropriate degree distribution.
Results on the eigenvalues of community matrix with the patch networks being Poisson networks are shown in~\cref{fig:poissonphaseplots,fig:LRevpoisson}.

\subsubsection{Networks where the giant component follows an exponential distribution}
\begin{figure}[t]
  \centering
  \includegraphics[width=.75\columnwidth]{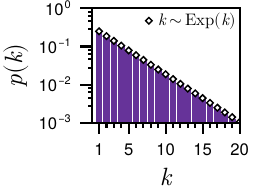}%
  \caption{%
    Degree distribution $10^4$ connected networks sampled using $p_0(k)$ given by \cref{eq:p0kconfig} and $p(k)$ an exponential distribution with $M=512$, $\langle k \rangle = 4$ and $k_{\min}=1$.
  Bars represent numerically obtained histogram and markers indicate the desired distribution given by~\cref{eq:SI:exp}.
  }%
  \label{fig:SI:pkExp}
\end{figure}

\label[supp]{sec:SI:exponential}%
Consider now a configuration model network whose giant component has an exponential degree distribution, $p(k,\alpha) = Ze^{- \alpha k}$, with $Z$ a normalization constant.
For $k_{\min} = 1$, it is convenient to write the degree distribution in terms of the mean degree $\langle k \rangle = 1/(1 - e^{-\alpha})$ as
\begin{align}
  \label{eq:SI:exp}
  p(k) = \frac{1}{\langle k \rangle}
  {\left(\frac{\langle k \rangle -1}{\langle k \rangle} \right)}^{k-1}
\end{align}
As with Poisson networks, we illustrate numerically obtained degree distributions with~\cref{eq:SI:exp} in~\cref{fig:SI:pkExp}, and the connected networks indeed have an exponential degree distribution as desired.
Results on the eigenvalues of a community matrix with the patch networks being exponential networks are shown in~\cref{fig:SI:LRevExp} (see also, \cref{sec:SI:stabilitysparse}).

\subsection{Sparse networks}\label[supp]{sec:SI:sparsenetworks}%
As posed in the main text (in particular, see~\cref{sec:topology} and~\cref{fig:poissonphaseplots}), we argue that ecological patch networks do not need to be dense in order for the system to be stable.
Here, we would like to formally define when we consider a network to be sparse, as network sparsity can be the result of different mechanisms.
The first, and perhaps most straightforward, mechanism the generates sparse networks is simply the absence of many edges.
For example, when considering Poisson networks (see~\cref{eq:SI:poisson}) --- or, more generally, Erd\H{o}s-R\'enyi networks --- the mean degree is given by
\begin{align*}
  \langle k \rangle = pM,
\end{align*}
where $p$ the probability of connecting two vertices.
Recall that we consider only connected networks, thus we need $p > \log M /M$.
In such networks, only when $p = \mathcal{O}(\log M / M)$, one could state that these networks are sparse, but obviously Erd\H{o}s-R\'enyi networks are not sparse for all $p$.
Here we say that only when the mean degree scales sub-linearly with the number of vertices, that is,
\begin{align}
  \label{eq:SI:sparsitydegree}
  \langle k \rangle \propto M^{\varphi - 1}, \qq{or} \langle k \rangle = o(M),
\end{align}
with $\varphi < 2$, one can formally define such networks as sparse\footnote{Note that most often $\varphi \geq 1$.}~\citep{newman2018networks,janson2018edge}.
One could further define ``truly sparse'' networks, for which $\langle k \rangle = \mathcal{O}(1)$, where often the degree distribution converges to a constant value independent of $M$.

Sparsity one can also be derived from the (ensemble averaged) number of edges, denoted with $E$, which can be expressed as a function of the mean degree,
\begin{align}
  E = \langle k \rangle M.
\end{align}
As the maximum total possible number of edges is $E_{\max} = \tfrac{1}{2}M(M-1)$, we obtain, for large $M$, that
\begin{align}
  \label{eq:SI:sparsityedges}
  E/E_{\max} \propto 1/M \rightarrow 0.
\end{align}
To summarize, one can verify whether~\cref{eq:SI:sparsitydegree} or~\cref{eq:SI:sparsityedges} hold and determine whether a specific network is dense, i.e.~$\langle k \rangle = \mathcal{O}(M)$, sparse, i.e.~$\langle k \rangle = o(M)$, or truly sparse, for which~$\langle k \rangle = \mathcal{O}(1)$.

\subsubsection{Stability in truly sparse networks}\label[supp]{sec:SI:stabilitysparse}%
\begin{figure}[t]
  \centering
  \includegraphics[width=.85\columnwidth]{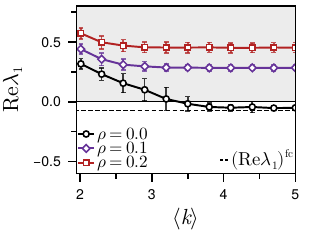}
  \caption{%
    Real part of the largest right-most eigenvalue versus the average degree $\langle k \rangle$ of community matrices for which the connected network has an exponential degree distribution.
    Note that, in contrast with Poisson networks (see~\cref{sec:SI:poisson} and~\cref{fig:poissonphaseplots,fig:LRevpoisson}), as $\langle k \rangle$ does not scale with $M$, these networks are sparse.
    Different interaction heterogeneities $\rho$ are shown. 
    Dashed line indicates approximated value in a fully connected network (shown here for $\rho=0$, see also~\cref{sec:SI:esd}).
    Relevant parameters are as in~\cref{fig:LRevpoisson}.
  }%
  \label{fig:SI:LRevExp}
\end{figure}
For configuration model networks with an exponential distribution (\cref{eq:SI:exp}), i.e.~exponential networks, the mean degree depends only on the parameter of the distribution, $\langle k \rangle = 1/(1-e^{-\alpha}) \equiv \mathcal{O}(1)$, and hence these networks are truly sparse by definition~\eqref{eq:SI:sparsitydegree}.
We here investigate the eigenvalues of communities matrices on top of these exponential networks.
Results are shown in~\cref{fig:SI:LRevExp} and show that $\lambda_1$ converges to approximately the same value of a fully-connected network as obtained using~\cref{eq:supportesdbaron}.
For our particular choice of values (see~\cref{fig:LRevpoisson}), we have that $E/E_{\max} \approx 0.01$ for $\langle k \rangle = 5$.
Therefore, with only approximately $1\%$ of the possible edges that could be present in the network, the conversion to the fully-connected case occurs before the graph could be considered dense.
The fact that convergence, which is accompanied by system stability, occurs in the sparse regime is critical, as it appears to indicate that ecological patch networks need not be densely connected to support stable meta-ecosystems.
As explained in the main text, the fact that stability of sparse networks is well-approximated by fully-connected ones could greatly simplify experimental validation, as one does not need to include (nearly) all possible dispersal pathways to emulate a fully-connected system.

\subsection{Small-world networks}\label[supp]{sec:SI:smallworld}%
As configuration network models are typically locally tree-like, meaning that short loops are absent~\citep{newman2009random}, the networks considered above do not display high degrees of clustering.
To this end, we consider the global clustering coefficient and the mean (shortest) path length.
The global clustering coefficient $C$ and the mean path length $\ell$ are defined as~\citep{newman2018networks}
\begin{align}
  \label{eq:SI:globalclusteringcoefficient}
  C &= \frac{3 \times \text{number of triangles}}{\text{number of all triplets}}, \\
  \ell &= \frac{1}{M(M-1)} \sum_{i\neq j} \Delta_{ij},
\end{align}
where $\Delta_{ij}$ is the shortest path length between nodes $i$ and $j$, thus $\ell$ is the average of the shortest path lengths between all pairs of nodes\footnote{Note that when $i$ and $j$ are not connected (i.e., there is no path between them), then typically $\ell = \infty$. However, as we consider only connected networks, the mean path lengths we consider are always finite.}.
In configuration network models, one finds that $C \sim \textrm{const.}/M$, and thus $C\rightarrow 0$ as $M$ grows~\citep{newman2001random}, and that, typically, $\ell$ is low relative to the size of the network.
\rev{That is, in sufficiently dense random networks one typically finds little clustering and low mean path lengths.}
However, in ecological networks (specifically spatial networks, see~\cref{sec:SI:spatialgraphs}), one could map the conceptualization of clustering in social networks to those of ecological nature.
In social networks, there is often a high probability that ``a friend of my friend is also my friend'' --- i.e.~a high tendency of triadic closure.
More formally, there is a high probability of an edge existing between two vertices that share a neighbor.
Within the context of dispersal, it is also reasonable to assume that when species can move to two adjacent patches, that these adjacent patches are also closeby, and thus a triangle is present (i.e., triadic closure).
To study the impact of clustering, we proceed by considering small-world networks, \rev{or (Newman-)Watts-Strogatz networks} that, over a wide range of parameters, display high degrees of clustering~\citep{watts1998collective}.
Note that we are aware that these are not considered to reflect real networks, see, e.g., \citep{newman2003properties}, but we use them purely as a means to isolate the effects of clustering on stability as much as possible.

\subsubsection{Watts-Strogatz networks}
\begin{figure}[t]
  \centering
  \includegraphics[width=\linewidth]{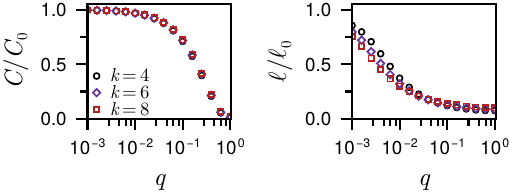}
  \caption{%
    Normalized global clustering coefficient $C/C_0$ and average path length $\ell/\ell_0$ in Watts-Strogatz networks with different edge densities $k$ for networks with $M=512$.
  }%
  \label{fig:SI:wsclusterpath}
\end{figure}

\label[supp]{sec:SI:smallworld:ws}
To generate Watts-Strogatz networks, one starts with a $k$-regular network wherein each vertex has exactly $k$ neighbors.
Then one iterates over all edges in the networks and rewires them randomly with probability $q$while avoiding self-loops and multiple edges.
When $q=0$ the network remains $k$-regular, while for $q=1$ we have a structure that resembles a random (Erd\H{o}s-R\'enyi) network.
Watts-Strogatz networks display the ``small-world phenomenon'', in that over a wide range of values for the rewiring probability $q$ high degrees of clustering and low average path lengths are obtained~\citep{watts1998collective} (see \cref{fig:SI:wsclusterpath}).
Finally, while Watts-Strogatz networks can, in principle, contain isolated vertices, we found that this rarely occurs in practice when $M$ is relatively large, and thus potential effects of this can be ignored safely.
\rev{To ensure that isolated subgraphs do not influence our results however, we only considered connected Watts-Strogatz networks in our analyses simply by omitting (i.e., resampling) those that are disconnected.}

\subsubsection{Newman-Watts-Strogatz networks}
\begin{figure}[t]
  \centering
  \includegraphics[width=\linewidth]{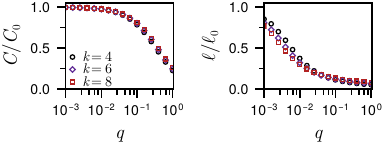}
  \caption{%
    Normalized global clustering coefficient $C/C_0$ and average path length $\ell/\ell_0$ in Newman-Watts-Strogatz networks with different initial edge densities $k$ for networks with $M=512$.
    Note the difference in the global clustering coefficient with Watts-Strogatz networks (\cref{fig:SI:wsclusterpath}), as even for $q=1$ the initial triangles remain, hence $C/C_0>0$.
  }%
  \label{fig:SI:nwsclusterpath}
  \includegraphics[width=.85\linewidth]{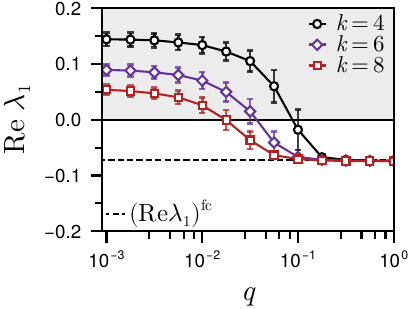}
  \caption{%
    Real part of the right-most eigenvalue versus edge adding probability $q$ of community matrices for which the network is a Newman-Watts-Strogatz network.
    Dashed line indicates approximated value in a fully connected network.
    Relevant parameters are as in~\cref{fig:LRevpoisson}, with $\rho=0$.
  }%
  \label{fig:SI:LRevNWS}%
\end{figure}

\label[supp]{sec:SI:smallworld:nws}
\rev{%
  A subtle but distinct variant of the Watts-Strogatz networks described above are Newman-Watts-Strogatz networks~\citep{newman1999scaling}.
  In Newman-Watts-Strogatz networks edges are added instead of being rewired.
  More specifically, during the iteration over all edges in the original $k$-regular network, a random edge is added with probability $q$.
  Note that this thus increases the density with increased $q$ however, and therefore thus does not isolate the effect of clustering.
  These networks also display the previously mentioned characteristics of small-world graphs (\cref{fig:SI:nwsclusterpath}).

  We additionally investigated stability in Newman-Watts-Strogatz networks, and results on the right-most eigenvalue are reported in~\cref{fig:SI:LRevNWS}.
  By comparing these results with those obtained in Watts-Strogatz networks (\cref{fig:LRevwattsstrogatz}), it is clear that adding edges, instead of rewiring, only has noticeable effect for lower edge densities $k$.
  Additionally, as was reported in~\cref{sec:topology}, triadic closure again appears to destabilize systems as high levels of clustering (obtained for low $q$) tend to lead to instability. 
}

\subsubsection{Disentangling clustering and path length: simulated annealing}\label[supp]{sec:SI:simanneal}%
\rev{%
  To disentangle the effects of clustering and path length further we would like to mention that we cannot rely on parametric network models to sample networks with arbitrary clustering and path lengths, as these, to the best of our knowledge, do simply not exist.
  We can however, slightly disentangle their effect on stability by studying dispersal networks for which the path length is altered using simulated annealing~\citep{reppas2015tuning}.
  
  While the search space for a standard multi-objective simulated annealing scheme is vast, we can rewire edges in such a way that the number of triangles, and thus the global clustering coefficient, remains the same while only the path length is affected.
  This thus transforms the multi-objective scheme, for clustering and path length, into a single-objective one only for path length.
  It can be implemented using a standard simulated annealing scheme with a slight adaptation in the procedure for choosing which edges to rewire.
  One starts with a random Watts-Strogatz network for some value of the rewiring probability $q$ and number of neighbors $k$.
  These initial networks have path length $\ell$ (\cref{fig:SI:wsclusterpath}.
  Networks are updated by first selecting two nodes, $v$ and $u$, which are (i)~not connected, (ii)~do not have any common neighbors.
  Then, for each node a random neighbor is selected, $v^\prime$ and $u^\prime$, for which (iii)~no common neighbor between them exists, and (iv)~both $v^\prime$ and $u^\prime$ do not form a triangle with any other neighbor of $v$ and $u$ respectively.
  Then, the network is rewired following a standard simulated annealing scheme.
  That is, rewire the network by connecting $v$ and $u$ and $v^\prime$ and $u^\prime$, compute the new path length $\ell^\prime$ if the network is still connected (otherwise the step is rejected), and evaluate the objective function $f(\ell) = ||\ell - \ell_{\textrm{goal}}||$.
  If $f(\ell^\prime) < f(\ell)$, or with probability $\exp \left( -\beta [f(\ell^\prime) - f(\ell)] \right)$, accept the new configuration or reject it.
  As is standard, the (inverse) temperature $\beta$ is reduced according to the desired annealing schedule.
  For more details on the implementation of this adapted simulated annealing scheme we refer the interested reader to Ref.~\citep{reppas2015tuning}.
  While this procedure does not allow one to tune $\ell$ arbitrarily, we have empirically verified that $\ell$ can be increased\footnote{Note that decreasing $\ell$ without changing $C$ is not possible as one will have to introduce triangles or introduce edges (instead of rewiring).} threefold within a reasonable number of steps when (see also~\citep[Fig.5]{reppas2015tuning}).
}

\begin{figure}[b]
  \centering
  \includegraphics[width=\linewidth]{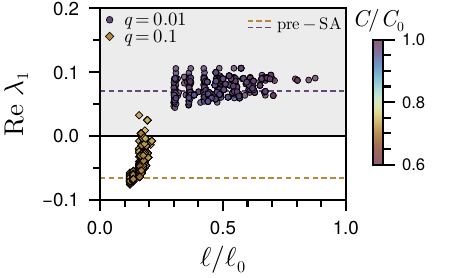}
  \caption{\rev{%
      Real part of the right-most eigenvalue versus mean path length for different two distinct levels of clustering (high, for $q=0.01$, and low, for $q=0.1$).
      Note that path length only affects stability when clustering is low.
      Networks before the simulated annealing (SA) scheme are Watts-Strogatz networks with rewiring probability $q$ and $k=6$.
      Dashed lines indicate $\Re\lambda_1$ before the SA scheme was applied, thus $C/C_0$ and $\ell/\ell_0$ are as in~\cref{fig:SI:wsclusterpath}.
      All other relevant parameters are as in~\cref{fig:LRevpoisson}, with $\rho=0$.
  }}%
  \label{fig:SI:simannealstability}
\end{figure}

\subsubsection{Disentangling clustering and path length: stability}\label[supp]{sec:SI:simannealstability}%
\rev{%
  Now that we can generate dispersal networks with some global clustering coefficient, and can then tune their mean path length, let us investigate the effect of path length on stability.
  For some values of the rewiring probability $q$ we have generated networks with distinct path lengths and have plotted the right-most eigenvalues in~\cref{fig:SI:simannealstability}.
  When clustering is high (low $q$), we see that systems are unstable over a vast range of path lengths.
  Only when clustering starts to decrease (high $q$) we find that path length starts to affect stability.
  These results appear to indicate that stability is more significantly influenced by triadic closure than by the mean path length.
  While triadic closure is most likely not the only (leading) network characteristic that determines stability, path lengths appear to only influence stability when clustering is (relatively) low.
  Yet, in those situations, the path length of the dispersal network does appear to influence stability significantly.
  A more in-depth investigation into possible underlying network characteristics that strongly affect stability is therefore justified, but is considered to be outside the scope of the analysis presented here.
}

\subsection{Spatial networks}\label[supp]{sec:SI:spatialgraphs}%
Spatial networks are networks whose vertices are explicitly embedded within a spatial domain \citep{herrmann2003connectivity}.
Within the context of ecology, they are a natural inclusion as real-world patch systems typically live on a two-dimensional plane \citep{fall2007spatial}.
Formally, spatial networks are defined by letting the vertices be distributed randomly in space with following some distribution $p(x)$. 
Vertices are then connected given some distance-related constraint, which can be interpreted as depending on the (typical) dispersal distance --- i.e.~the dispersal kernel \citep{clobert2012dispersal,grilli2015metapopulation} --- of the species considered. 
Thus, given two vertices $\mu$ and $\nu$, located at $\vb*{x}^\mu$ and $\vb*{x}^\nu$ respectively, they are connected by an edge if
\begin{align}
  \label{eq:SI:distance}
  || \vb*{x}^\mu - \vb*{x}^\nu || \leq \theta,
\end{align}
where $\theta$ is the threshold, cutoff, or the typical scale of connections.
It is convenient to further let define a constant connectivity $\kappa$ that ensures well-defined degree distributions when $M\rightarrow \infty$.
Note that for this one has to take $\theta \rightarrow 0$, thus the connectivity is defined as
\begin{align}
  \label{eq:SI:spatialconnectivity}
  \kappa = MV(\theta),
\end{align}
where $V(\theta)$ is the volume of the ball around a vertex \citep{dall2002random,herrmann2003connectivity}.
Note that the connectivity $\kappa$ further fixes the mean degree as
\begin{align}
  \label{eq:SI:spatialmeandegree}
  \langle k \rangle = \kappa \int dx \; p^2(x).
\end{align}
Moreover, one can derive the degree distribution of a spatial network with spatial distribution $p(x)$, which reads \citep{herrmann2003connectivity}
\begin{align}
  \label{eq:SI:spatialdegreedistribution}
  p(k;\kappa) = \frac{\kappa^k}{k!} \int dx \; p^{k+1}(x) e^{-\kappa p(x)}.
\end{align}

\begin{figure}[t]
  \centering
  \includegraphics[width=.8\columnwidth]{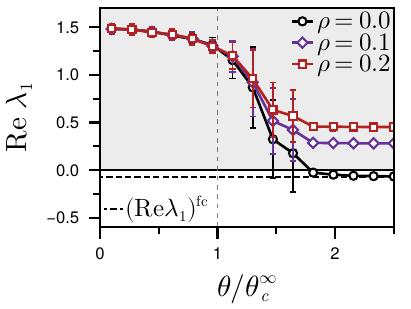}%
  \caption{%
    Real part of the right-most eigenvalues for two-dimensional random geometric graphs for different interaction heterogeneities $\rho$.
    Patches (vertices) are connected when their distance is less than $\theta$ (~\cref{eq:SI:distance}).
    Network connectivity (average degree, see text) is defined by~$\kappa = \pi M \theta^2$, and the connectance threshold for $M\rightarrow \infty$ is $\kappa^\infty_c \approx 4.5$ in two-dimensional networks~\citep{dall2002random}.
    Dashed line indicates approximated value in a fully connected network (shown here for $\rho=0$, see also~\cref{sec:SI:esd}).
    Relevant parameters are as in~\cref{fig:LRevpoisson}.
  }%
  \label{fig:SI:LRevrgg}
\end{figure}

\subsubsection{Random geometric graphs}\label[supp]{sec:SI:rgg}%
When the spatial distribution $p(x)$ is uniform, the resulting networks are called \emph{random geometric graphs} \citep{dall2002random}.
It further simplifies~\cref{eq:SI:spatialmeandegree,eq:SI:spatialdegreedistribution}, meaning that random geometric graphs are networks with mean degree $s=\kappa$ and a degree distribution that is Poissonian.
However, it should be noted that random geometric graphs are different from the random configuration model networks with Poissonian degree distributions we discussed earlier in~\cref{sec:SI:configmodel} (see Ref.~\citep[pp. 8-9]{dall2002random}).
More specifically, the degree distribution does not uniquely define a network (or an ensemble of networks), as different network characteristics, such as the clustering coefficient, depend strongly on the process by which the network is generated~\citep{falkenberg2020identifying}.

Note that, despite random geometric graphs being an excellent example of spatial networks that have been studied extensively in ecological literature~\citep{grilli2015metapopulation,barter2017spatial}, their clustering coefficient depends only on the embedded dimension, and not on the threshold considered~\citep{dall2002random}.
As such, they might not accurately model patterns observed in real-world connected patch systems.
To include further structure, one obvious way is to change the spatial distribution $p(x)$ to reflect the spatial characteristics of interest~\citep{herrmann2003connectivity}, or to consider a more realistic growth process~\citep{plaszczynski2022levy}.
However, these extensions are considered to be out of the scope of the work presented here.

\subsubsection{Random geometric graphs and isolated nodes}%
\begin{figure}[h]
  \centering
  \includegraphics[width=.7\columnwidth]{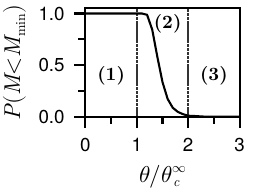}
  \caption{%
    Probability of cluster sizes not exceeding the minimum required size $M_{\min}$ in random geometric graphs, where $M_{\min}$ can be derived from the stability criterion in fully-connected networks (\cref{eq:Mmin}).
    Other relevant parameters for computing the minimum size, here $M_{\min} \approx 100$, are $S=100$, $M=512$, $c=0.2$, $\sigma\sqrt{cS}=1$, and $b - r = \sigma\sqrt{cS/M}$.
  }%
  \label{fig:SI:probcluster}
\end{figure}

To the best of our knowledge, there is currently no available method to construct random geometric graphs that are connected regardless of the choice of $\theta$.
In fact, because random geometric graphs are spatially embedded, connected networks should arise only when $\theta$ is large enough.
In order to study the effects of dispersal on stability in these networks however, it is necessary to see whether small isolated clusters do not persist when $\theta \gg \theta^\infty_c$.
To this end, we computed the probability of an isolated cluster to be smaller than the minimum required size $M_{\min}$ (see~\cref{eq:Mmin}).
The results, shown in~\cref{fig:SI:probcluster}, indicate that clusters of size $M < M_{\min}$ indeed do not persist when $\theta$ is large enough.
As such, the results presented in~\cref{fig:rggphaseplot} hold, as there are no isolated nodes that render the systems unstable by virtue of being unstable themselves (as discussed in~\cref{sec:SI:isolated}).

\end{document}